\documentclass[12pt,oneside,reqno]{amsart}

\newtheorem{theorem}{Theorem}[section]

\theoremstyle{definition}

\theoremstyle{remark}
\newtheorem{remark}[theorem]{Remark}

\numberwithin{equation}{section}

\usepackage{amsmath,amsthm,amsfonts,amssymb}
\usepackage[margin=1 in]{geometry}
\usepackage{comment}
\usepackage{caption}
\usepackage{graphicx}
\usepackage{epstopdf}
\usepackage{subfigure}
\usepackage{natbib}
\usepackage{tikz}
\usepackage{bbm}
\usepackage{times}
\usepackage{algorithmic}
\usepackage{algorithm}  
\usepackage[T1]{fontenc}
\usepackage{color}

\makeatletter
\def\listofalgorithms{\@starttoc{loa}\listalgorithmname}
\def\l@algorithm{\@tocline{0}{3pt plus2pt}{0pt}{1.9em}{}}
\makeatother

\DeclareMathAlphabet{\MATHIT}{OT1}{ptm}{m}{it}
\DeclareSymbolFont{Letters}{OML}{ztmcm}{m}{it}
\DeclareSymbolFontAlphabet{\mathNormal}{Letters}

\floatname{algorithm}{Algorithm}

\begin{document}

\title[Numerical Scheme for  Investment-Consumption Under Proportional Transaction Cost]{A Numerical Scheme for A Singular control problem: Investment-Consumption Under Proportional Transaction Costs}

\author{Wan-Yu Tsai}
\address { Wan-Yu Tsai, Arash Fahim \newline
	\indent Department of Mathematics \newline
	\indent Florida State University \newline
	\indent 1017 Academic Way \newline
	\indent Tallahassee, FL-32306 \newline
	\indent United States of America}
\email{wtsai@math.fsu.edu, fahim@math.fsu.edu}

\author{Arash Fahim}
\thanks{A. Fahim is partially supported by the NSF (DMS-1209519).}
\date{\today}

\subjclass[2010]{} 

\begin{abstract}
This paper concerns the numerical solution of a fully nonlinear parabolic double obstacle problem arising from a finite portfolio selection with proportional transaction costs. We consider optimal allocation of wealth among multiple stocks and a bank account in order to maximize the finite horizon discounted utility of consumption. The problem is mainly governed by a time-dependent  Hamilton-Jacobi-Bellman equation with gradient constraints. We propose a numerical method which is composed of Monte Carlo simulation to take advantage of the high-dimensional properties and finite difference method to approximate the gradients of the value function. Numerical results illustrate behaviors of the optimal trading strategies and also satisfy all qualitative properties proved in \cite{dai09} and \cite{chen2013characterization}.     
\end{abstract}

\maketitle

\smallskip
\noindent \textbf{Keywords:} Hamilton-Jacobi-Bellman equation, stochastic control, Monte Carlo approximation, backward stochastic differential equations, portfolio optimization, transaction costs.

\section{Introduction}
This paper presents the numerical solution of an optimal investment-consumption problem in the presence of proportional transaction costs during a finite time period. Given a known initial wealth, the objective of an investor is to decide the best consumption and investment strategy which maximizes the expected discounted utility of consumption over the finite investment period. In the absence of transaction costs and for specific utility functions, the solution can be exactly obtained and an investor's optimal trading strategy is to maintain a constant proportion of wealth invested in risky stocks, which is called the \textit{Merton proportion} shown by \cite{merton71}. This constant proportion depends on the investor's risk preference and also the market parameters. Merton's strategy, simply stated, is to continuously rebalance portfolio holdings in order to keep the fraction of investment in risky assets constant. However, in the presence of transaction costs, a continuous portfolio rebalancing process may incur infinite costs. Thus, the question arises: \textit{what is the optimal strategy if there are transaction costs in the market? }

Transaction cost appears in different ways, as a fixed commission or a proportion to the size of trade. This paper deals with the case where there is only proportional transaction costs; for a review of constant cost or a mixture of both, see \cite{ARS16} and references therein. \cite{MC76} are the first to introduce proportional transaction costs into Merton's model. They provide a valuable insight on the optimal strategy; i.e. an investor should maintain the fraction of wealth in risky assets inside a so-called {\it no-trading region} and trading only takes place along the boundary of the no-trading region. As a consequence, the crucial question is: \textit{how to identify the optimal no-trading region which corresponds to the optimal trading strategy?}

Under certain restricted settings, this question has been partially answered. When the market is confined to consist of a single risky asset and a bank account, \cite{davis90} give a rigorous analysis of the classical Merton's problem with proportional transaction costs over infinite time horizon. The optimal policy is formulated as a nonlinear free boundary problem which separates the buying and the selling regions from the no-trading one. Their paper contains detailed characterization, both theoretical and numerical, of the value function and optimal policies under certain assumptions. \cite{SS94} relax assumptions of \cite{davis90}'s problem, and apply the viscosity solution approach to provide regularity and existence results. Many other papers have carried out an asymptotic analysis including \cite{JS04}, \cite{GO10}, and \cite{KM10}. A thorough convergence proof for general utility functions is studied by \cite{ ST13}, and an extension to several risky assets is considered by \cite{PST12}. Other numerical schemes have been proposed by \cite{TZ94} and \cite{TZ97} for general utility functions, and by \cite{muthuraman2006multidimensional} for a model with more than one risky asset. Nevertheless, these papers only deal with the infinite horizon scenario where the no-trading region does not evolve in time, and are based on finite difference/element method which are not efficient in higher dimensions.

Theoretical analysis on the finite-time problem has been studied recently and is restricted to the no consumption case with a single risky asset. \cite{liu04} first shows analytical properties of the optimal investment problem with a deterministic finite horizon. \cite{dai2009finite} establish a link between the singular control problem and the obstacle problem, and completely characterize the behaviors of the resulting free boundaries. Numerical solution of this optimal investment problem is proposed by \cite{arregui2012numerical}. More recently, there is a plethora of literature devoted to the characterization of optimal investment-consumption strategy. \cite{dai09} consider the investment and consumption optimization decision in finite time horizon, and characterize the behaviors of free boundaries for a single risky asset case. \cite{dai08} propose the penalty method to demonstrate the numerical solution to a singular control problem arising from portfolio selection with proportional transaction costs.  \cite{B12} provides a proof to the same problem with power utility function by expanding the value function into a power series, and obtains a ``nearly optimal'' strategy.  

In the present paper, we propose a numerical scheme based on Monte Carlo simulation for the optimal investment-consumption problem with proportional transaction costs and deterministic time horizon. As discussed in the next section, the value function of such control problem is characterized by a  \textit{Hamilton-Jacobi-Bellman (HJB) equation}. The existing numerical schemes for this HJB equation in the literature including \cite{TZ94}, \cite{TZ97} and \cite{muthuraman2006multidimensional} are based on finite difference/element method, which are only practical in low dimensional problems.  Moreover, the dimension can be higher in many applications, especially in finance problems. Thus, we propose a numerical technique that combines Monte Carlo simulation with finite difference discretization so as to solve the nonlinear double obstacle problem, and aim to characterize the free boundaries and qualitative properties of the solution.

Our numerical scheme is strongly motivated by the aforementioned work of \cite{fahim2011probabilistic} who introduce the backward probabilistic numerical scheme combined with Monte Carlo and finite difference method for high-dimensional fully nonlinear partial differential equations. They decompose the scheme into two steps. First, the Monte Carlo step includes isolating the linear generator of some underlying diffusion process to split the PDE into this linear part and a remaining nonlinear one. Then, a projection method is employed to evaluate the remaining nonlinear part of the PDE. In this paper, we will modify the numerical method to incorporate the free boundaries on the no-trading region. Moreover, we will show that the proposed method can work in the case of correlated stocks. It is worth noticing that the type of free boundaries in this current problem is different from the obstacle problem such as the one in \cite{BF14} and therefore the scheme developed in this paper is not in the same nature of Monte Carlo scheme. We believe the motivation behind this proposed method can be extended to various HJB for singular control problems.

This paper is organized as follows. In Section 2 the optimal investment and consumption problem with proportional transaction costs is presented. Section 3 is dedicated to some simplifications of the control problem in Section 2. The numerical scheme composed of Monte Carlo simulation and finite difference discretization is proposed in Section 4. In Section 5, we show that the implementation of the proposed numerical scheme is compatible with the theoretical results in \cite{dai09} and \cite{chen2013characterization} in a single risky asset or two risky assets cases. Several examples that illustrate performances of the proposed numerical method are also presented in this section. And the last section draws some conclusion.

\section{The optimal investment-consumption problem}
We consider an optimal investment-consumption problem in finite time horizon $T \in (0, \infty)$ with proportional transaction costs, the model being the same in \cite{dai08} and \cite{chen2013characterization}. 

Suppose a continuous time market consisting of one risk-free asset and multiple risky assets available for investment. The risk-free asset (bank account), denoted by $S^{0}_{t}$, pays an interest rate $r>0$ continuously and thus can be expressed as
\begin{equation}
dS_{t}^{0} = rS_{t}^{0}dt.
\end{equation}  
Let $N$ be the number of available risky investments, called ``stocks'' hereafter. The $N$ stocks have constant mean rates of return $\alpha_{1}, \alpha_{2}, \cdots, \alpha_{N}$. We denote the vector of $N$ stock prices by $S_{t}=(S^{1}_{t}, S^{2}_{t}, \cdots, S^{N}_{t})^{'}$ and the mean rates of return by $\alpha=(\alpha_{1}, \alpha_{2}, \cdots, \alpha_{N})^{'}$. The evolution of stocks can be written as
\begin{equation}
dS_{t} = {\rm\bf diag}(S_{t})(\alpha dt+\sigma dB_{t}),
\end{equation}
where ${\rm\bf diag}(S_{t})$ is the $N \times N$ matrix formed with elements of $S_{t}$ as its diagonal, $\sigma$ denotes the $N \times N$ positive definite covariance structure, and $\{B_{t}: t \in [0,T]\}$ is a standard $N$-dimensional Brownian motion defined on a filtered probability space $(\Omega, \mathcal{F}, \{\mathcal{F}_{t}\}_{0 \leq t \leq T},\mathbb{P})$.

Assume that an investor holds a portfolio $(X_{t}, Y_{t})'=(X_{t}, Y^{1}_{t}, \cdots, Y^{N}_{t})'$, where $X_{t}$ and $Y^{i}_{t}$ are dollar amount invested in the bank account and in the $i^{th}$ stock at time $t$. His problem is to choose a consumption and investment strategy over the deterministic horizon in order to maximize his objective: the discounted utility of consumption during the investment period. We require that the consumption $c_t$ must be non-negative and occur from cash in the bank, and its process $c_t$ should be adapted to $\mathcal{F}_{t}$ and integrable for any finite $t$, that is,
\begin{equation}
\int_{0}^{t} c_{s}ds <\infty  \quad  \forall t \geq 0.
\end{equation}
Now we introduce two $\mathcal{F}_{t}$-adapted processes $L_{t}=(L^{1}_{t}, \cdots, L^{N}_{t})'$ and $M_{t}=(M^{1}_{t}, \cdots, M^{N}_{t})'$ which are non-negative, non-decreasing, and right continuous with left limits (RCLL). $L^{i}_{t}$ represents the cumulative dollar value spent for the purchase of stock $i$ before incurring transaction costs, whereas $M^{i}_{t}$ represents the cumulative amount of money obtained from the sale of stock $i$. Denote the transaction costs for buying and selling stocks by $\lambda=(\lambda_{1}, \lambda_{2}, \cdots, \lambda_{N})' \geq 0$ and $\mu=(\mu_{1}, \mu_{2}, \cdots, \mu_{N})'\geq 0$ respectively. To be more precise, buying a unit of stock $i$ will cost $(1+\lambda_{i})$ in cash from the bank and selling a unit of stock $i$ will receive $(1-\mu_{i})$ in cash added into the bank. We assume that $\lambda_{i}+\mu_{i}>0, i=1,2, \cdots, N$. With transaction costs and consumption, the controlled evolution of $X_{t}$ and  $Y_{t}$ can be described by the following equations
\begin{eqnarray}
dX_{t}&=&(rX_{t}-c_{t})dt-(e+\lambda) \cdot dL_{t}+(e-\mu) \cdot dM_{t},\\
dY_{t}&=& {\rm\bf diag}(Y_{t})\left[\alpha dt +\sigma dB_{t}\right]+dL_{t}-dM_{t}.
\end{eqnarray}
Here, ``$\cdot$'' is the standard dot product and $e$ is a vector of ones with appropriate length. 

We require the investor's net wealth  at any time to be positive because he would not be bankrupt if he is forced to liquidate his position. If taking transaction costs into consideration, the investor's net wealth at time $t$ is given by $X_{t}+\sum_{i=1}^{N} \min\left[(1+\lambda_{i})Y^{i}_{t},(1-\mu_{i})Y^{i}_{t}\right]$. Therefore, we define the \textit{solvency region} $\mathcal{S_{\lambda, \mu}}$ as
\begin{equation}
\mathcal{S_{\lambda, \mu}} = \left\{ (x,y)\in (\mathbb{R}, \mathbb{R}^{N}) : x+\sum_{i=1}^{N} \min\left[(1+\lambda_{i})y_{i},(1-\mu_{i})y_{i}\right]\geq 0 \right\}.
\end{equation}
Given an initial position $(X_{0},Y_{0})'=(x,y)' \in \mathcal{S_{\lambda, \mu}}$, an investment-consumption strategy $(c_{t}, L_{t}, M_{t})$ is called admissible if and only if the portfolio position $(X_{t}, Y_{t})$ lies in $\mathcal{S_{\lambda, \mu}}$ for all $t \in [0,T)$. Let $\mathcal{A}_{t}(x,y)$ be the set of admissible strategies. The investor's objective consists of choosing an admissible strategy so as to maximize the expected discounted utility of accumulative consumption and the terminal wealth, that is,
\begin{equation}
\sup_{(c_{t},L_{t},M_{t})\in \mathcal{A}_{0}(x,y)} \mathbb{E}_{0}^{x,y} \left[\int_{0}^{T} e^{-\beta t}U(c_{t})dt+e^{-\beta T} U(W_{T}) \right],
\end{equation}
where $\beta > 0$ is the discount factor, $\mathbb{E}_{t}^{x,y}$ denotes the conditional expectation at time $t$ given an initial endowment $X_{t}=x$, $Y_{t}=y$, $W_{T}$ is the terminal net wealth given by $W_{T} = X_{T} + \sum_{i=1}^{N} \min \left[ (1+\lambda_{i})Y^{i}_{T}, (1-\mu_{i})Y^{i}_{T} \right]$, and $U$ is the utility function which belongs to the class of \textit{ Constant Relative Risk Aversion} (CRRA) utility functions, i.e.
\begin{equation}
U(c) = \begin{cases} \frac{c^\gamma}{\gamma} & \mbox{if } \gamma < 1, \gamma \neq 0, 
\bigskip
\\ \log (c) & \mbox{if } \gamma =0. \end{cases}
\end{equation}
Here $\gamma$ is the relative risk aversion coefficient that describes the investor's risk preference. These utility functions are well-known and have been used very wildly in modelling the risk preference of an investor. Then we define the value function by
\begin{eqnarray}
V(x,y,t)&=&\sup_{(c_{t},L_{t},M_{t})\in \mathcal{A}_{t}(x,y)} \mathbb{E}_{t}^{x,y} \left[\int_{t}^{T} e^{-\beta (s-t)}U(c_{s})ds+e^{-\beta (T-t)} U(W_{T}) \right],
\label{value function}
\end{eqnarray}
for $(x, y) \in \mathcal{S_{\lambda, \mu}}, \; t \in [0,T)$.

\section{The HJB equation and scaling}
{\color{black} By applying the dynamic programming arguments [cf. Section IV.3, \cite{fleming2006controlled}], the value function $V$ of the stochastic control problem  (\ref{value function}) satisfies the following \textit{Hamilton-Jacobi-Bellman} (HJB) equation:}

\begin{eqnarray}
0=\min \bigg \{ -\partial _{t} V-\frac{1}{2} \sum_{i,j=1}^{N} a_{ij} y_{i}y_{j}\partial _{y_{i}y_{j}}V - \sum_{i=1}^{N} \alpha_{i}y_{i}\partial _{y_{i}}V - rx\partial _{x}V + \beta V - U^{*}(\partial _{x}V),  \nonumber\\
\min_{i} \big[ -(1-\mu_{i})\partial _{x} V+ \partial _{y_{i}}V \big], \quad \min_{i} \big[(1+\lambda_{i})\partial _{x}V-\partial _{y_{i}}V\big] \bigg \},
\label{HJB}
\end{eqnarray}
with the terminal condition
\begin{equation}\label{eqn:terminal}
V(x,y,T) = U \left( x+ \sum_{i=1}^{N} \min \big[ (1+\lambda_{i})y_{i}, (1-\mu_{i})y_{i} \big] \right), 
\end{equation}
where $a=[a_{ij}]_{i,j=1}^N= \sigma \sigma^{'}$ and 
\begin{equation*}
U^{*}(\nu) = \sup_{c\geq 0}\left\{ U(c)-c\nu \right\}  = \begin{cases} \frac{1-\gamma}{\gamma} (\nu) ^{\frac{\gamma}{\gamma-1}} & \mbox{if } \gamma < 1, \gamma \neq 0, 
\bigskip
\\ -\log (\nu)-1 & \mbox{if } \gamma =0. \end{cases}
\end{equation*}
In this paper, we focus on the computational scheme to solve equation (\ref{HJB}) with terminal condition \eqref{eqn:terminal}. 
\begin{remark}
Equation \eqref{HJB} can be interpreted in the variational inequality sense, i.e.
\begin{enumerate}
	\item The value function $V$ satisfies all three following inequalities
	\begin{equation*}
	\begin{split}
		0\le& -\partial _{t} V-\frac{1}{2} \sum_{i,j=1}^{N} a_{ij} y_{i}y_{j}\partial _{y_{i}y_{j}}V - \sum_{i=1}^{N} \alpha_{i}y_{i}\partial _{y_{i}}V - rx\partial _{x}V + \beta V - U^{*}(\partial _{x}V),\\
		0\le& \min_{i} \big[ -(1-\mu_{i})\partial _{x} V+ \partial _{y_{i}}V \big],\\ 
		0\le&\min_{i} \big[(1+\lambda_{i})\partial _{x}V-\partial _{y_{i}}V\big].
	\end{split}
	\end{equation*}
	\item 
	If $ \displaystyle 0< \min_{i} \big[ -(1-\mu_{i})\partial _{x} V+ \partial _{y_{i}}V \big]$ and  $ \displaystyle 0< \min_{i} \big[(1+\lambda_{i})\partial _{x}V-\partial _{y_{i}}V\big]$, we must have
	\begin{equation*}
	0= -\partial _{t} V-\frac{1}{2} \sum_{i,j=1}^{N} a_{ij} y_{i}y_{j}\partial _{y_{i}y_{j}}V - \sum_{i=1}^{N} \alpha_{i}y_{i}\partial _{y_{i}}V - rx\partial _{x}V + \beta V - U^{*}(\partial _{x}V)
	\end{equation*}
\end{enumerate}
\end{remark}

Following \cite{dai08}, we use the homothetic property of the value function to reduce the dimensionality of the problem for further numerical analysis. For any constant $\rho>0$, the ``homothetic property'' of the value function is as follows:
\begin{eqnarray}
V(\rho x, \rho y, t) = \begin{cases} \rho^{\gamma}V(x,y,t) & \mbox{if } \gamma < 1, \gamma \neq 0, 
\bigskip
\\ \big(\frac{1-e^{-\beta (T-t)}}{\beta}+e^{-\beta (T-t)}\big)\log (\rho)+V(x,y,t) & \mbox{if } \gamma =0. \end{cases} \nonumber
\end{eqnarray}
This property allows us to reduce the dimension of the original problem from $N+1$ to $N$ by adopting the wealth fraction as state variables. Indeed, we define a new function
\begin{equation}
\varphi(y,t)=V(1-e \cdot y, y, t),
\label{newvaluefunction}
\end{equation}
where $e$ is a vector of ones with length $N$, and $y$ represents the vector of the fraction of wealth invested in each stock when the total wealth $w$ is one ($w=1$). It is clearly sufficient to compute $\varphi(y,t)$ since the original value function is then given by $$V(x,y,t)=\varphi\left(\frac{y}{x+e \cdot y},t\right)(x+e \cdot y)^{\gamma}.$$ 

{\color{black}The derivation of the HJB equation and the computational procedure for both the log utility and the power utility functions are the same. Therefore we provide a detailed description of the power utility case only.} In terms of $\varphi(y,t)$, the HJB equation in (\ref{HJB}) for the power utility function ($U(c)=c^\gamma / \gamma$) becomes
\begin{eqnarray}
0=\min \left\{ -\partial _{t} \varphi-\mathcal{\hat{L}}\varphi, \; \min_{i} \hat{S}_{i} \varphi, \; \min_{i} \hat{B}_{i} \varphi \right\}, \label{HJB system}
\end{eqnarray}
with the terminal condition
\begin{eqnarray}
\varphi(y,T) = \gamma^{-1}{\Big( 1+\sum_{i=1}^{N} \min \left\{ -\mu_{i}y_{i}, \lambda_{i}y_{i} \right\} \Big)^{\gamma}} \quad \text{ for } y \in  \Theta ^{N}, \nonumber
\end{eqnarray}
where $$\Theta ^{N} = \{ (y_{1}, y_{2}, \cdots, y_{N})\in \mathbb{R}^{N}: 1+\sum_{i=1}^{N} \min \left\{ -\mu_{i}y_{i}, \lambda_{i}y_{i}\right\} \geq 0  \},$$
and
\begin{eqnarray}
\mathcal{\hat{L}}\varphi &=& \frac{1}{2} \sum_{i,j=1}^{N}  \eta_{ij} \partial _{y_{i}y_{j}} \varphi + \sum_{i=1}^{N} b_{i} \partial _{y_{i}} \varphi -\vartheta \varphi + \frac{1-\gamma}{\gamma} \left( \gamma \varphi - \sum_{i=1}^{N} y_{i} \partial _{y_{i} }\varphi \right)^{\frac{\gamma}{\gamma -1}}, \label{eq:notrade}\\
\hat{S}_{i} \varphi &=& \left[ \mu_{i} \gamma \varphi - \sum_{k=1}^{N}(-\delta_{ik}+\mu_{i}y_{k})\partial _{y_{k}}\varphi \right], \label{eq:sell}\\
\hat{B}_{i} \varphi &=& \left[ \lambda_{i} \gamma \varphi - \sum_{k=1}^{N} (\delta_{ik}+\lambda_{i}y_{k})\partial _{y_{k}}\varphi\right], \label{eq:buy}
\end{eqnarray}
with
\begin{eqnarray}
\eta_{ij} &=& y_{i}y_{j}\sum_{k=1}^{N}\sum_{\ell=1}^{N} a_{k\ell}(\delta_{ik}-y_{k})(\delta_{j\ell}-y_{\ell}), \label{eta}\\
b_{i} &=& \frac{1}{2} \sum_{k=1}^{N}\sum_{\ell=1}^{N} a_{k\ell}y_{k}y_{\ell}(\gamma -1)(\delta_{ik}+\delta_{i\ell}-2y_{i})+\sum_{k=1}^{N} y_{k}(\delta_{ik}-y_{i})(\alpha_{k}-r), \label{b}\\
\vartheta &=& \beta -\gamma \left( r + \frac{1}{2} \sum_{k=1}^{N}\sum_{\ell=1}^{N} a_{k\ell}y_{k}y_{\ell}(\gamma -1) + \sum_{k=1}^{N} (\alpha _{k}-r)y_{k}  \right). \label{vartha} 
\end{eqnarray}
Here $\delta _{ij}$ represents the Kroneker index with $\delta _{ij} = 1$ if $i=j$ and $\delta _{ij} = 0$ otherwise. 
The above dimension reduction technique has been wildly used; see, for example, \cite{dai09} for $N=1$, and \cite{muthuraman2006multidimensional} for $N=2$ without the time variable. 

\begin{figure}[!t]
	\subfigure[Homotheticity property]{\label{fig:a}\includegraphics[width=0.43\linewidth]{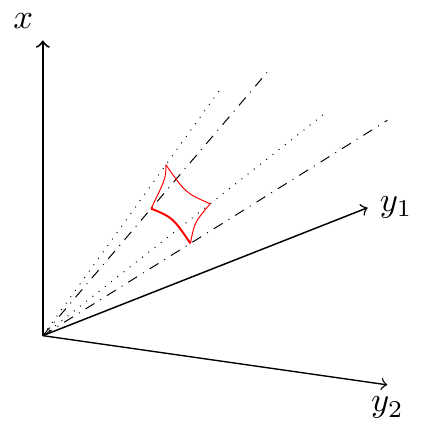}}
	\hfill
	\subfigure[$\Theta^{2}$ region and Trading strategy ]{\label{fig:b}\includegraphics[width=0.48\linewidth]{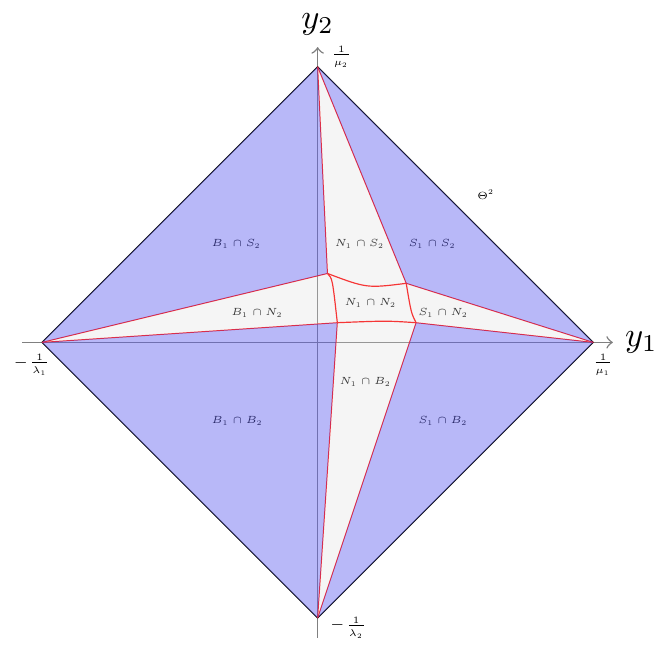}}
	
	\caption{Trading and no-trading regions along the $w=1$ cut at time $t$}
	\label{fig:image2}
\end{figure}

We consider a portfolio which consists of a risk-free asset and two risky assets ($N=2$) for illustration purpose. Before adopting the homothetic property, this problem is three dimensions, and the polygon cone in Figure \ref{fig:a} is the no-trading region. When we apply the homothetic property and rewrite the value function $V(x,y,t)$ by $\varphi(y,t)$ defined in (\ref{newvaluefunction}), we can reduce the problem to two dimensions. The red region shown in Figure \ref{fig:a} represents the no-trading region with the wealth equals one cut at time $t$ after dimension reduction.

Now we define the following representations for later use. Let
\begin{eqnarray}
B_{i} &=& \{ (y,t) \in \Theta^{N} \times [0,T):  \hat{B}_{i} \varphi = 0 \}, \\
S_{i} &=& \{ (y,t) \in \Theta^{N} \times [0,T): \hat{S}_{i}  \varphi = 0\}, \\
N_{i} &=& \Theta^{N} \times [0,T) \setminus (B_{i} \cup S_{i}),
\end{eqnarray}
where $B_{i}$, $S_{i}$ and $N_{i}$ represent the buying region, selling region, and no-trading region with respect to the $i^{th}$ stock. For illustration, we consider the case of $N=2$ as well. Figure \ref{fig:b} shows that the domain $\Theta^{2}$ at time $t$ is partitioned into different regions along the wealth equals one cut. In the area filled with gray, the investor should buy or sell one stock just enough to push the fraction back to the no-trading region $N_{1} \cap N_{2}$. In the area filled with blue, it is not possible to trade only one stock to make the fraction reaching the no-trading region. Thus, two stocks should be transacted simultaneously to reach the corner of inaction region $N_{1} \cap N_{2}$.

Although we can formulate the value function as the HJB equation presented in (\ref{HJB system}), the complete analytical solution cannot be obtained. Also, the standard numerical methods, such as finite difference method and finite element method, work only for low dimensional cases. Due to the curse of dimensionality and the lack of an exact solution, the development of appropriate numerical methods are highly desirable to approximate the solution and provide qualitative properties under different model parameter settings. Therefore, in the next section we propose a numerical scheme that combines Monte Carlo simulation with finite difference discretization in order to solve this nonlinear variational inequality problem.

\section{Numerical Method}
As mentioned before, the main goal of this paper is to propose an appropriate numerical method in order to approximate the solution of the optimal investment-consumption problem and consequently obtain the trading strategies. We first notice that the main difficulties associated with the numerical solution of the HJB in (\ref{HJB system}) are twofold:

\begin{enumerate}
	\item the free boundary feature related to the double obstacle problem,
	\item the equation presents a nonlinear term in (\ref{eq:notrade}), $\left( \gamma \varphi - \sum_{i=1}^{N} y_{i} \partial _{y_{i} }\varphi \right)^{\frac{\gamma}{\gamma -1}}$.
\end{enumerate}

\subsection{The two-step procedure}
In order to overcome the free boundary feature, we will use a two-step procedure which extends the idea proposed by \cite{muthuraman2006multidimensional}. Step 1 solves the nonlinear second order PDE in the first part of the HJB equation while step 2 updates the value function in different regions of the domain $\Theta^{N}$. To be more precise, we begin by finding the values $\varphi (y,t)$  for all $y$ in the domain $\Theta^{N}$ such that the first part of HJB equation in (\ref{HJB system}) holds true, that is,
\begin{equation}
-\partial_{t}\varphi - \mathcal{\hat{L}}\varphi = 0  \quad \text{for} \quad y \in \Theta^{N}, \label{eq:nonlinear}
\end{equation}
{\color{black}with the boundary condition
\begin{equation*}
\varphi (y,t)=
\begin{cases} 0 & \mbox{if } \gamma > 0, \\ 
-\infty & \mbox{if } \gamma <0, \end{cases} \quad \text{for} \quad y \in \partial \Theta^{N}, \quad t \in [0,T). 
\end{equation*} 
The reason that we set the boundary condition in (\ref{eq:nonlinear}) to be zero or negative infinity is because on the boundary of solvency region, the investor is forced to liquidate his position at any time $t$ and his net wealth on the boundary is zero. Since we consider the power utility function, the utility of consumption is zero for $\gamma >0$ and negative infinity for $\gamma <0$ because of zero net wealth. For the power utility function, we can set $$\varphi (y,t) = \begin{cases} 0 & \mbox{if } \gamma > 0, \\ 
-\infty & \mbox{if } \gamma <0, \end{cases}$$ for $y \in \partial \Theta^N \cup (\mathbb{R}^N \setminus \Theta^N)$, $t \in [0,T)$ because the investor's position never exits the solvency region. This will later become useful in the numerical implementation where we have to define the value function at the discrete points outside the region $\Theta^{N}$.}

We also require that the other two formulas in (\ref{HJB system}) should be satisfied in the domain $\Theta^{N}$. Hence, in the next step our procedure deals with the free boundary terms $\hat{S}_{i}\varphi$ and $\hat{B}_{i}\varphi$ in (\ref{eq:sell}) and (\ref{eq:buy}) for all $i=1,2,\cdots,N$. We have to find the point $y^*$ where $\hat{S}_{i}\varphi (y^*,t)$ and/or $\hat{B}_{i}\varphi (y^*,t)$ are negative, and then adjust the value at the point such that the variational inequalities hold true. Denote the trading strategy for buying and selling stocks at $(y^*,t)$ by $$\upsilon^{(y^*,t)} = (\mathbbm{1}_{\{\hat{B}_{1}(y^*,t) < 0\}} , \cdots, \mathbbm{1}_{\{\hat{B}_{N}(y^*,t) < 0\}})',$$ and $$\varrho^{(y^*,t)} = (\mathbbm{1}_{\{\hat{S}_{1}(y^*,t) < 0\}} , \cdots, \mathbbm{1}_{\{\hat{S}_{N}(y^*,t) < 0\}})',$$ where $\mathbbm{1}$ is an indicator function. We update the value function at $(y^*,t)$ by
\begin{equation}
\varphi (y^*,t) = \varphi (\bar{y},t) \left( \frac{1+ \sum_{i=1}^{N} \lambda _{i} y^*_{i} \upsilon _{i}^{(y^*,t)} - \sum_{i=1}^{N} \mu _{i} y^*_{i} \varrho_{i}^{(y^*,t)}}{1+ \sum_{i=1}^{N} \lambda _{i} \bar{y_{i}} \upsilon _{i}^{(y^*,t)}  - \sum_{i=1}^{N} \mu _{i} \bar{y_{i}} \varrho _{i}^{(y^*,t)} } \right)^{\gamma}.
\label{update}
\end{equation}
Here $\bar{y}$ is the point that satisfies the following conditions:
\begin{enumerate}
    \item $\bar{y}$ is the point that is closest to $y^*$ along the characteristic curves in the region which includes $y^*$,
    \item $\bar{y}$ is the point on the boundary of the no-trading region facing the region to which $y^*$ belongs.
\end{enumerate}
\noindent This procedure will be repeated  backward in time until a sequence of no-trading regions and trading strategies are obtained at each time step.

\begin{figure}[!t]
	\includegraphics[width=0.63\linewidth, height=0.58\linewidth]{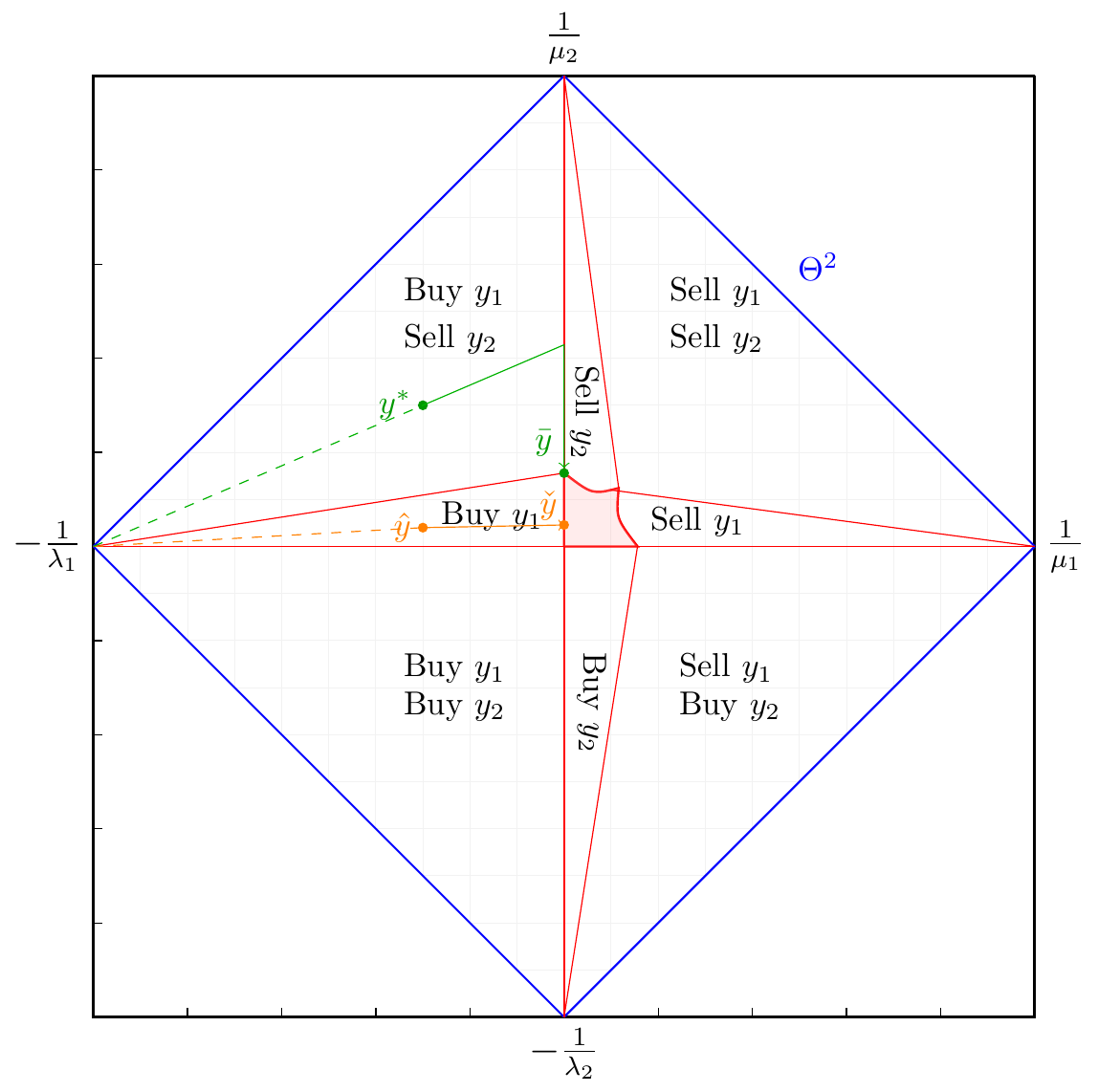}
	
	\caption{Illustration for the two-step procedure with $N=2$ at time $t$}
	\label{fig:image3}
\end{figure}

For illustration purpose, we use Figure \ref{fig:image3} to convey the idea of the two-step procedure when $N=2$ at a given time $t$. The region inside the blue diamond refers to the domain $\Theta^{2}$. First, we have to solve $\varphi (y,t)$ for $y \in \Theta^{2}$ satisfying the nonlinear second order partial differential equation in (\ref{eq:nonlinear}). Once the values $\varphi (y,t)$ are known, we check the gradient constraints $\hat{S}_{i}\varphi$ and $\hat{B}_{i}\varphi$ for $i=1,2$. The no-trading region filled with red meets the conditions $\hat{S}_{i}\varphi \geq 0$ and $\hat{B}_{i}\varphi \geq 0$ for $i=1,2$, and thus the value $\varphi (y,t)$ in this region do not need to be changed. However, for example, if $y^*=(y^*_1,y^*_2)$ marked green in Figure \ref{fig:image3} is the point such that  $\hat{B}_{1}\varphi (y^*,t)<0$ and $\hat{S}_{2}\varphi (y^*,t)<0$, $y^*$ is classified as an element in the set $B_{1} \cap S_{2}$ and also its value $\varphi (y^*,t)$ should be adjusted to meet the conditions $\hat{B}_{1}\varphi (y^*,t)=0$ and $\hat{S}_{2}\varphi (y^*,t)=0$. In Figure \ref{fig:image3}, $\bar{y}=(\bar{y_1},\bar{y_2})$ marked green is the point which is closest to the point $y^*$ along the characteristic curves and also on the boundary of the no-trading region facing the region to which $y^*$ belongs. Therefore we update the value function by
$$ \varphi (y^*,t) = \varphi (\bar{y},t) \left( \frac{1+\lambda_{1} y^*_1- \mu_{2} y^*_2}{1+\lambda_{1}\bar{y_1}- \mu_{2} \bar{y_2}}\right)^{\gamma}.  $$
It means that the investor should buy the first stock and sell the second one to rebalance his position so as to reach the corner of no-trading region. For another example, $\hat{y}=(\hat{y}_1,\hat{y}_2)$ marked orange in Figure \ref{fig:image3} is the point such that  $\hat{B}_{1}\varphi (\hat{y},t)<0$, and then we have to adjust its value function by $$ \varphi (\hat{y},t) = \varphi (\check{y},t) \left( \frac{1+\lambda_{1} \hat{y}_1}{1+\lambda_{1}\check{y}_1}\right)^{\gamma}, $$
where $\check{y}$ is the point marked orange in Figure \ref{fig:image3} that is closest to $\hat{y}$ along the characteristic curve in the $B_{1}$ region.

\subsection{The computational scheme for solving the nonlinear second order PDE} Due to the curse of dimensionality and the lack of an analytical solution, an appropriate numerical method for solving high-dimensional fully nonlinear PDEs is highly desirable. Our numerical method is mainly motivated by the recent work of \cite{fahim2011probabilistic}. The computational scheme they provided consists of two parts. First, the Monte Carlo step includes isolating the linear generator of some underlying diffusion process to split the PDE into this linear part and a remaining nonlinear one.  Next, discrete-time finite difference approximation is applied to evaluate the remaining nonlinear part of the PDE along the underlying diffusion process. The first part takes the advantage of the high-dimensional property of Monte Carlo method while the second part deals with the nonlinear term of the equation. In this paper, we modify the numerical method to incorporate the free boundaries on the no-trading region.

\subsubsection{\textbf{Notation}}
We shall first introduce some notations. The collection of $n \times d$ matrices with real entries is denoted by $\mathcal{M}(n,d)$. For a matrix $A \in \mathcal{M}(n,d)$, $A'$ represents its transpose and $\sqrt{A}$ returns square root of each element in the matrix. For $A, B \in \mathcal{M}(n,d)$, we define $A \cdot B := Tr[A'B]$. In particular, $A$ and $B$ are vectors of $\mathbb{R}^{n}$ when $d=1$ and $A \cdot B$ reduces to the standard dot product. $D$ and $D^2$ are the gradient and the Hessian matrix defined by 
\begin{eqnarray}
D \varphi=\Big(\frac{\partial \varphi}{\partial y_{1}}, \frac{\partial \varphi}{\partial y_{2}}, \cdots, \frac{\partial \varphi}{\partial y_{N}}\Big)'  \qquad \text{ and }  \qquad D^2 \varphi = \left[ \arraycolsep=0.5pt\def\arraystretch{1.8}
\begin{array}{@{}cccc@{}}
\frac{\partial ^2 \varphi}{\partial y_{1}^{2}} & \frac{\partial ^2 \varphi}{\partial y_{1} \partial y_{2}} & \cdots & \frac{\partial ^2 \varphi}{\partial y_{1} \partial y_{N}}\\
\frac{\partial ^2 \varphi}{\partial y_{2} \partial y_{1}} & \frac{\partial ^2 \varphi}{\partial y_{2}^{2}} & \cdots & \frac{\partial ^2 \varphi}{\partial y_{2} \partial y_{N}}\\
\vdots & \vdots & \ddots & \vdots \\
\frac{\partial ^2 \varphi}{\partial y_{N} \partial y_{1}} & \frac{\partial ^2 \varphi}{\partial y_{N} \partial y_{2}} & \cdots & \frac{\partial ^2 \varphi}{\partial y_{N}^{2}}
\end{array}\right].
\label{derivative}
\end{eqnarray}

Let $b=(b_1, b_2, \cdots, b_N)'$ be a vector of $\mathbb{R}^N$ where $b_{i}$ is the coefficient of the first derivative of $\varphi$ with respect to the variable $y_i$ in (\ref{b}), and $\eta \in \mathcal{M}(N,N)$ be a matrix with elements of $\eta_{ij}$ at row $i$ and column $j$ in (\ref{eta}). The diagonal matrix $ \xi \in \mathcal{M}(N,N)$ is defined by $\xi:={\rm\bf diag}(\eta)$. Next, we determine the linear operator
\begin{equation}
L^{\hat{Y}} \varphi := \partial _{t} \varphi + b \cdot D\varphi + \frac{1}{2} \; \xi \cdot D^2 \varphi. \nonumber
\end{equation}
Then the remaining nonlinear parts are represented as
\begin{equation}
F(y,t, \varphi , D\varphi , D^2 \varphi) := \frac{1}{2} \sum_{i,j=1, j\neq i}^{N} \eta_{ij} \partial _{y_{i}y_{j}} \varphi - \vartheta \varphi + \frac{1-\gamma}{\gamma} \left( \gamma \varphi - \sum_{i=1}^{N} y_{i} \partial _{y_{i} }\varphi \right)^{\frac{\gamma}{\gamma -1}}. \nonumber
\end{equation}
Hence, the problem we have to deal with becomes
\begin{eqnarray}
0 &=& -L^{\hat{Y}} \varphi (y,t, \varphi , D\varphi , D^2 \varphi) - F(y,t, \varphi , D\varphi , D^2 \varphi) \; \text{for} \; y \in \Theta ^{N}, \; t=[0,T); \label{eq:MonteCarloSystem}\\
\nonumber \\
\varphi(y,T) &=& \gamma^{-1}{\Big( 1+\sum_{i=1}^{N} \min \left\{ -\mu_{i}y_{i}, \lambda_{i}y_{i} \right\} \Big)^{\gamma}} \quad \text{ for } y \in \Theta ^{N}. \label{terminal}
\end{eqnarray}

\subsubsection{\textbf{Discretization}}
As with any numerical scheme, the first step is to discretize the time space and the domain of state variables. Let $h:=T/n$ be the time step, and $t_{k}=kh$, $k=0,1,\cdots,n$ for a positive integer $n$. Suppose we have a uniform grid, denoted by $G^{\; t_{k}}_{\Delta y}$, for the domain $\Theta ^{N}$ with the grid size $\Delta y = (\Delta y_{1}, \Delta y_{2},\cdots, \Delta y_{N})'$ in each state variable direction. Denote a discretized point with $y=(y_{1},y_{2},\cdots,y_{N})' \in G^{\; t_{k}}_{\Delta y}$ at time $t_{k}$ by $(y,t_{k})$.

Let $B_{t}$ be an $\mathbb{R}^{N}$-dimensional standard Brownian motion defined in section 2. Consider the one-step-ahead Euler discretization of the diffusion $\hat{Y}$ corresponding to the linear operator $L^{\hat{Y}}$
\begin{equation}
\hat{Y}^{y,t_{k+1}}_{h}:=y+b(y,t_{k})h+\sqrt{\xi (y,t_{k})}\;(B_{t_{k+1}}-B_{t_{k}}). \label{eq:Euler}
\end{equation}
If we assume that the nonlinear PDE in (\ref{eq:nonlinear}) has a solution, we follow from It\^{o}'s formula and replace the process $\hat{Y}$ by its Euler discretization to get
\begin{equation}
\mathbb{E}_{t_{k},y} \left[ \varphi (\hat{Y}^{y,t_{k+1}}_{h}, t_{k+1}) \right] = \varphi  (y, t_{k}) + \mathbb{E}_{t_{k},y} \left[ \int_{t_{k}}^{t_{k+1}} L^{\hat{Y}} \varphi (\hat{Y}_{s},s, \varphi , D\varphi , D^2 \varphi) \; ds  \right],
\end{equation}
where $\mathbb{E}_{t_{k},y}:=\mathbb{E}[\cdot | \hat{Y}_{t_{k}}=y]$ is the conditional expectation, and $D^{\kappa}$ is the $\kappa^{th}$ order partial differential operator with respect to the space variable $y$ defined in (\ref{derivative}). By approximating the integral, the value function $\varphi (y, t_{k})$ can be evaluated as follows:
\begin{eqnarray}
\varphi (y, t_{k}) &=& \mathbb{E}_{t_{k},y} \left[ \varphi  (\hat{Y}^{y,t_{k+1}}_{h}, t_{k+1}) \right] -  h \; L^{\hat{Y}} \varphi (y, t_{k}, \mathcal{D}^0 \varphi, \mathcal{D}^1 \varphi, \mathcal{D}^2\varphi) + \mathcal{O}(h), \label{eq:iter0}\\
\mathcal{D}^{\kappa}\varphi &:=& \mathbb{E}_{t_{k},y}[D^{\kappa} \varphi  (\hat{Y}^{y,t_{k+1}}_{h}, t_{k+1})], \quad \kappa=0,1,2. \label{eq:iter3}
\end{eqnarray}
Since $\varphi$ is also a solution to the PDE in  (\ref{eq:MonteCarloSystem}) which means $$L^{\hat{Y}} \varphi  (y, t_{k}, \mathcal{D}^0 \varphi, \mathcal{D}^1 \varphi, \mathcal{D}^2\varphi) = - F(y, t_{k}, \mathcal{D}^0 \varphi, \mathcal{D}^1 \varphi, \mathcal{D}^2\varphi),$$ we have the discretized approximation of the value function as follows:
\begin{equation}
\varphi^{h}(y,t_{n}) := \gamma^{-1}\Big( 1+\sum_{i=1}^{N} \min \left\{ -\mu_{i}y_{i}, \lambda_{i}y_{i} \right\} \Big)^{\gamma} \quad \text{ for } y \in G^{\; t_{n}}_{\Delta y}, \label{eq:iter1}
\end{equation}
and for $y \in G^{\; t_{k}}_{\Delta y}$, $ k=0,\cdots,n-1$
\begin{equation}
\varphi^{h}(y,t_{k}) := \mathbb{E}_{t_{k},y} \left[ \varphi ^{h}(\hat{Y}^{y,t_{k+1}}_{h}, t_{k+1}) \right] + h \; F(y, t_{k}, \mathcal{D}^0 \varphi^{h}, \mathcal{D}^1 \varphi^{h}, \mathcal{D}^2\varphi^{h}). \label{eq:iter2}
\end{equation}

Once the linear operator $L^{\hat{Y}} \varphi$ is chosen, the remaining nonlinear parts are handled by means of classical centered difference approximation. Let $e_{i}$ be the unit vector in the $y_{i}$ direction and then the first order term $\partial_{y_{i}} \varphi ^{h} (y,t_{k})$ is discretized by the centered difference approximation of the gradient, that is,
\begin{equation}
\partial_{y_{i}} \varphi ^{h}(y,t_{k}) \approx \frac{\varphi^{h} (y+ \Delta y_{i} e_{i},t_{k}) -  \varphi ^{h}(y-\Delta y_{i}e_{i},t_{k})}{2 \Delta y_{i}}. \label{eq:firstderivative}
\end{equation}
The cross derivative term $\partial_{y_{i}y_{j}} \varphi^{h}$ is discretized as follows
\begin{equation}
\begin{split}
\partial_{y_{i}y_{j}} \varphi^{h} (y,t_{k}) \approx \frac{1}{4\Delta y_{i} \Delta y_{j}} & \Biggr[  \varphi^{h} (y+ \Delta y_{i} e_{i}+\Delta y_{j} e_{j},t_{k})+\varphi ^{h}(y- \Delta y_{i} e_{i}-\Delta y_{j} e_{j},t_{k}) \\
&- \varphi ^{h}(y+ \Delta y_{i} e_{i}-\Delta y_{j} e_{j},t_{k})-\varphi^{h}(y- \Delta y_{i} e_{i}+\Delta y_{j} e_{j},t_{k}) \Biggr]. \label{eq:secondderivative}
\end{split}
\end{equation}

\begin{algorithm}[!t]
	\caption{Mixed Monte Carlo Simulation and Finite Difference Method Algorithm}
	\label{alg:Framwork}
	\begin{algorithmic}[1]
		\ENSURE{The value function $\varphi ^{h}(y,t)$, and the optimal buying and selling boundaries}
		
		\bigskip
		\bigskip
		
		\STATE{Let $h:=T/n$ and $t_{k}=kh$, $k=0,1,\cdots,n$}  be the time step
		\STATE{Discretize the domain $\Theta ^{N}$ into uniform grid, denoted by $G^{\; t_{k}}_{\Delta y}$, with the grid size $\Delta y = (\Delta y_{1}, \Delta y_{2},\cdots, \Delta y_{N})'$ in each state variable direction}
		\FOR{ each $y \in G^{\; t_{n}}_{\Delta y}$}
		\STATE{Set the value function $\varphi^{h} (y,t_{n})$ at time $t_{n}$ according to its terminal condition in (\ref{eq:iter1})}
		\STATE{Evaluate $\partial_{y_{i}} \varphi^{h} (y,t_{n})$ and $\partial_{y_{i}y_{j}} \varphi^{h} (y,t_{n})$ for $i,j=1,2,\cdots,N$ in each state variable direction by centered-difference approximation in (\ref{eq:firstderivative}) and (\ref{eq:secondderivative})}
		\ENDFOR
		
		\FOR{$\ell=n-1 \; $; $\; \ell \geq 0 \;$; $\; \ell=\ell -1$}
		\FOR{ each $y \in G^{\; t_{\ell}}_{\Delta y}$}
		\STATE{Generate $M$ sample paths of $\hat{Y}^{y,t_{\ell +1}}_{h}$ by the one-step-ahead Euler discretization in (\ref{eq:Euler}})
		\STATE{Estimate the values $\varphi^{h} (\hat{Y}^{y,t_{\ell +1}}_{h}, t_{\ell +1})$, $\partial_{y_{i}}\varphi^{h} (\hat{Y}^{y,t_{\ell +1}}_{h}, t_{\ell+1})$, and $\partial_{y_{i}y_{j}}\varphi^{h} (\hat{Y}^{y,t_{\ell +1}}_{h}, t_{\ell+1})$ by linear interpolation if the simulated point $\hat{Y}^{y,t_{\ell+1}}_{h}$ is not on the grid}
		\STATE{Approximate $\mathcal{D}^{\kappa}\varphi^{h}$ for $\kappa=0,1,2$ in (\ref{eq:iter3}) by $\mathbb{\hat{E}}^{M}[D^{\kappa} \varphi^{h} (\hat{Y}^{y,t_{\ell +1}}_{h}, t_{\ell+1})]$ corresponding to the sample size $M$}
		\STATE{Compute $\varphi^{h} (y,t_{\ell})$ based on  (\ref{eq:iter2})}
		\ENDFOR
		\FOR{ each $y \in G^{\; t_{\ell}}_{\Delta y}$}
		\STATE{Find the grid point $y^*$ where $\hat{S}_{i}\varphi ^{h}(y^*,t_{\ell}) <0$ and/or $\hat{B}_{i}\varphi ^{h}(y^*,t_{\ell}) <0$  for $i=1,2,\cdots,N$ in (\ref{eq:sell}) and (\ref{eq:buy}), and then adjust the value $\varphi ^{h}(y^*,t_{\ell})$ by (\ref{update}) such that the $\hat{S}_{i}\varphi ^{h}(y^*,t_{\ell})=0$ and/or $\hat{B}_{i}\varphi ^{h}(y^*,t_{\ell})=0$} 
			
		\ENDFOR
		\ENDFOR
	\end{algorithmic}	
\end{algorithm}

Once we have the set of one-step-ahead random path simulations $\hat{Y}^{y,t_{k+1}}_{h}$, the iteration computes the discrete solution $\varphi ^{h}(y, t_{k})$ at time $t_{k}$ from $\varphi ^{h}(y, t_{k+1})$ by (\ref{eq:iter3})-(\ref{eq:iter2}). Note that if the simulated point $\hat{Y}^{y,t_{k+1}}_{h}$ is not on the grid $G^{\; t_{k}}_{\Delta y}$, we will approximate the value $\varphi ^{h} (\hat{Y}^{y,t_{k+1}}_{h}, t_{k+1})$, $\partial_{y_{i}}\varphi ^{h}(\hat{Y}^{y,t_{k+1}}_{h}, t_{k+1})$, and $\partial_{y_{i}y_{j}}\varphi ^{h} (\hat{Y}^{y,t_{k+1}}_{h}, t_{k+1})$ by interpolation. The interpolated value at a query point is based on linear interpolation of the values at neighboring grid points in each respective dimension.

In view of the above interpretation associated with the value function, our numerical scheme studied in this paper can be expressed as a mixed Monte Carlo simulation and finite difference method. The Monte Carlo portion includes the choice of an underlying diffusion process while the finite difference portion consists of the derivative approximation of the remaining nonlinearity. We summarize the two-step iterative procedure in Algorithm \ref{alg:Framwork}.

\begin{remark} 
The numerical method in Algorithm \ref{alg:Framwork} is inspired by \cite{fahim2011probabilistic} where they developed a Monte Carlo scheme for fully nonlinear PDEs of the form
$$
\begin{cases}
0=\mathcal{L}v+  G(y,t, v , Dv , D^2 v)\\
v(T,y)=g(y)
\end{cases}
$$
where $\mathcal{L}$ is a linear parabolic operator and $G$ is a nonlinear parabolic operator.
In the numerical scheme of \cite{fahim2011probabilistic}, they use the linear parabolic operator $\mathcal{L}$ to generate sample paths of the diffusion process.
Therefore, one has some flexibility in choosing the underlying diffusion process of the samples paths; e.g. one can also choose a linear parabolic operator $\mathcal{L}_1$  to generate the diffusion sample paths as long as the nonlinear term
\[
F(y,t, \varphi , D\varphi , D^2 \varphi):=(\mathcal{L}-\mathcal{L}_1)\varphi+G(y,t, \varphi , D\varphi , D^2 \varphi),
\]
remains parabolic. $G(y,t,r,p,\gamma):\mathbb{R}^N\times[0,T]\times\mathbb{R}\times\mathbb{R}^n\times\mathcal{M}(N,N)\to\mathbb{R}$ is called parabolic if $\nabla_\gamma G$ is positive definite  where $\nabla$ denotes the vector differential operator.

The numerical scheme in Algorithm
\ref{alg:Framwork} sets
\[
L^{\hat{Y}} \varphi :=\mathcal{L}_1\varphi=\partial _{t} \varphi + b \cdot D\varphi + \frac{1}{2} \; {\rm\bf diag}(\eta) \cdot D^2 \varphi
\]
and leaves the off-diagonal terms
\[
\frac{1}{2} \sum_{i,j=1, j\neq i}^{N} \eta_{ij} \partial _{y_{i}y_{j}} \varphi
\]
for the nonlinear part. It is simply because the diffusion process simulated by this parabolic operator is less complicated when we have only diagonal elements. On the other hand, inclusion of off-diagonal  second order derivative terms does not affect the sufficient conditions in \cite{fahim2011probabilistic} for the convergence of the numerical scheme, i.e. consistency, stability and monotonicity. For instance, monotonicity in \cite{fahim2011probabilistic} is guaranteed by the assumption that ${\rm Tr}[{\rm\bf diag}(\eta)^{-1}\nabla_\gamma F]\le 1$; that is, the diffusion coefficient in $\mathcal{L}_1$  dominates the derivative of the nonlinear operator $F$ with respect to the component of $D^2v$. Since $\mathcal{L}_1$ has the diagonal elements of the second order derivative and $\mathcal{L}-\mathcal{L}_1$ has only the off-diagonal elements, we have ${\rm Tr}[{\rm\bf diag}(\eta)^{-1}\nabla_\gamma F]={\rm Tr}[{\rm\bf diag}(\eta)^{-1}\nabla_\gamma G]$. 

It is worth mentioning that the adjustment in Step 15 of Algorithm \ref{alg:Framwork} to handle the free boundary makes it difficult to show the scheme is monotone. Therefore, we restrict our study to the numerical convergence of the proposed scheme.
\end{remark}

\section{Numerical Results}
The objectives in this section are: (1) to examine the performance of the mixed Monte Carlo simulation and finite difference method algorithm applying on the investment-consumption optimization problem; (2) to indicate the behaviors of optimal trading strategies.

\subsection{Test 1}

In this first example, we consider the following set of financial parameters:
\begin{equation*}
N=1, \quad r=0.07, \quad \alpha_{1}=0.12, \quad \sigma_{11}=0.4, \quad \beta=0.1, \quad \gamma=0.2, \quad \mu_{1}=\lambda_{1}=0.05,
\end{equation*} 
and solve the problem for the time interval $t \in [0,5]$. The theoretical properties of the solution and free boundaries to the problem (\ref{HJB system}) for $N=1$ case are presented in \cite{dai09}. We will use the following proven statements to verify the numerical results obtained from our mixed Monte Carlo / finite difference method.

Let \begin{equation}
\tau = \frac{1}{\alpha_{1}-r}\log\Big(\frac{1+\lambda_{1}}{1-\mu_{1}}\Big) \quad \text{and} \quad \tilde{y}=-\frac{\alpha_{1}-r-(1-\gamma)\sigma_{11}^{2}}{\alpha_{1}-r}.
\end{equation}

\noindent According to Theorem 5.4 in \cite{dai09}, the two free boundaries $B_{t}$ and $S_{t}$ for the $N=1$ case should satisfy the following properties:
\begin{enumerate}
		\item for $t \in [0, T)$,  $$B_{t} < S_{t}, \text{ and } S_{t} \geq S_{T^{-}}=\frac{1}{1+(1-\mu_{1})\tilde{y}};$$
		moreover, $$S_{t}=1  \quad \text{if} \quad \alpha_{1}-r-(1-\gamma)\sigma_{11}^{2} =0,$$
		$$S_{t}>1  \quad \text{if} \quad \alpha_{1}-r-(1-\gamma)\sigma_{11}^{2} >0,$$
		$$S_{t}<1  \quad \text{if} \quad \alpha_{1}-r-(1-\gamma)\sigma_{11}^{2} <0.$$
		\item for $t \in [0, T)$, $$B_{t} \leq \frac{1}{1+(1+\lambda_{1})\tilde{y}},$$
		and $$B_{t}=0  \quad \text{if and only if} \quad t \in [T-\tau, T).$$
\end{enumerate}

\noindent Now we have the values $\alpha_{1}-r-(1-\gamma)\sigma_{11}^{2} \approx -0.078 <0$  and $\tau \approx 2.002$ so that the selling and buying boundaries should satisfy
\begin{equation*}
\begin{cases} 
\displaystyle {\frac{1}{1+(1-\mu_{1})\tilde{y}} \approx 0.4029  \leq S_{t}<1} & \mbox{for } t \in [0,T); \bigskip\\

\displaystyle{B_{t} \leq \frac{1}{1+(1+\lambda_{1})\tilde{y}} \approx 0.3791}  & \mbox{for }t \in [0, T-\tau); \text{ and } B_{t}=0  \text{ for } t \in [T-\tau, T).
\end{cases}
\end{equation*}

\begin{figure}[!t]
	\subfigure[$\alpha_{1}=0.12$]{\label{fig:image4}\includegraphics[width=0.49\linewidth]{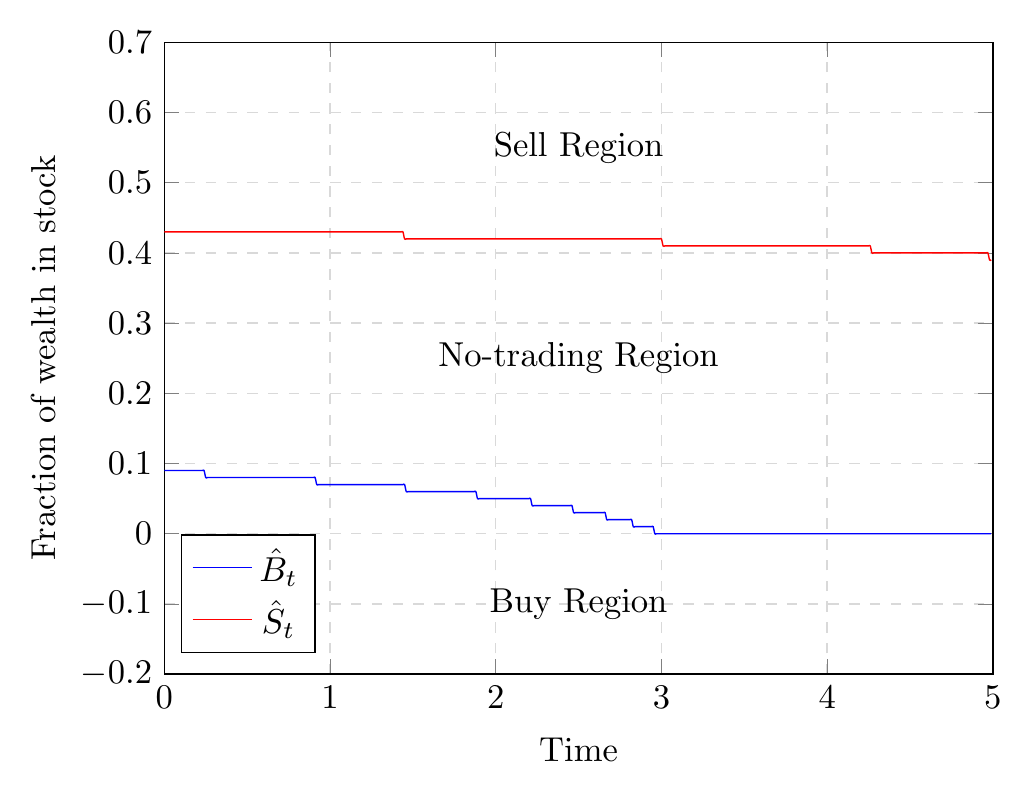}}
	\subfigure[Varying $\alpha_{1}$]{\label{fig:image5}\includegraphics[width=0.49\linewidth]{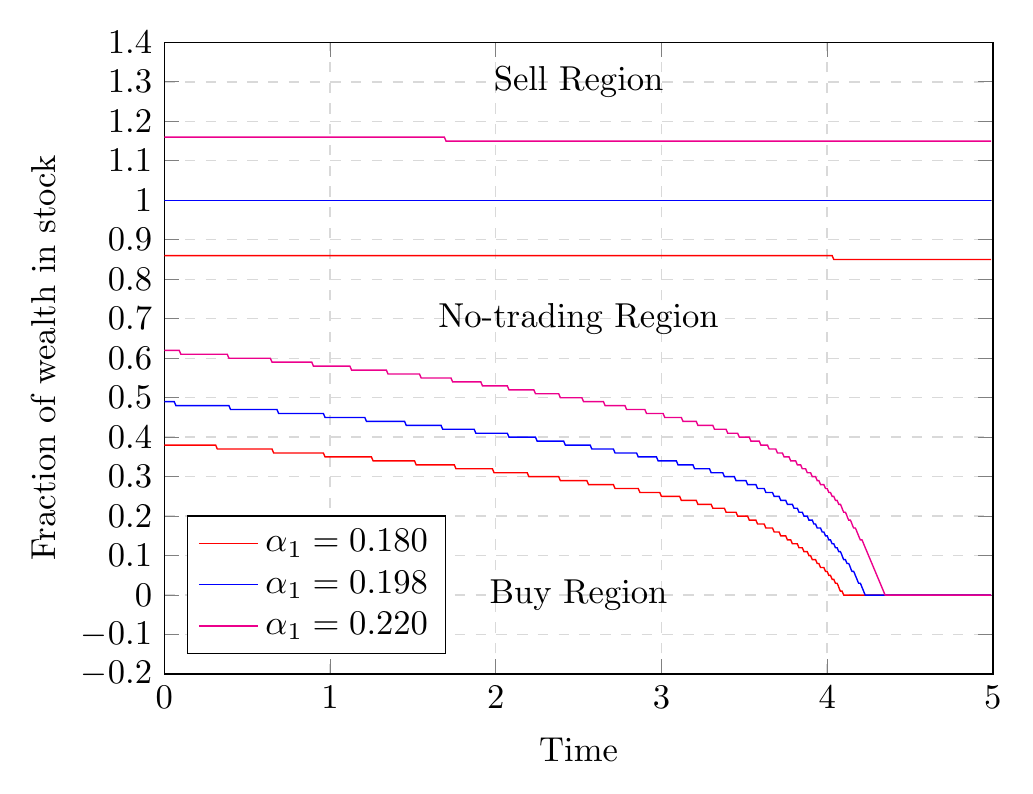}}
	
	\caption{The estimated selling, buying, and no-trading regions for the $N=1$ case. Test 1 parameters: $ \; r=0.07, \; \sigma_{11}=0.4,\; \mu_{1}=\lambda_{1}=0.05, \; \gamma=0.2, \; \beta=0.1$}
\end{figure}

Concerning the numerical method, we use the time step $h=0.01$ and uniform grid with length $\Delta y = 0.01$. Another numerical parameter that we have used is the number of simulated sample paths $M=10^{5}$.  Figure \ref{fig:image4} shows the numerical approximation of the optimal trading strategies in the fraction of wealth in stock at each time step. The upper function is the selling boundary while the lower one is the buying boundary. Clearly, these two boundaries depend on time, and the no-trading region is between these two boundaries. First, we have verified that the theoretical properties are satisfied for all the grid points at every discrete time step. Also, it shows that the value of the buying boundary tends to zero as the time is greater than $T-\tau \approx 3$ which indicates that it is suboptimal to buy a risky asset soon as the finite horizon is approaching. This phenomenon, known as ``no-buying near maturity'', was first proved by \cite{dai09} with consumption and transaction costs in finite time horizon. Furthermore, the optimal selling boundary is always greater than the buying one which mainly points out that a risk averse investor prefers to buy low and sell high.

Figure \ref{fig:image5} shows the optimal trading boundaries with varying $\alpha_{1}$. We can observe that both the buying and selling boundaries increase as the value of $\alpha_{1}$ increases, which indicates that the investor should hold a larger fraction of wealth in risky asset when the return of risky asset is higher. If $\alpha_{1}<0.198$, the selling boundary is less than one which means it is always suboptimal to leverage. However, leverage will be needed if $\alpha_{1}>0.198$. The obtained numerical results are again in full agreement with the theoretical properties stated in \cite{dai09} Theorem 5.4.

\subsection{Test 2}

In this second numerical test, the following financial parameter values have been considered:

\begin{eqnarray*}
&& N=2, \quad r=0, \quad \beta=0.1, \quad \gamma=0.2, \quad \mu_{1}=\lambda_{1}=\mu_{2}=\lambda_{2}=0.05 \\
&& \alpha_{1}=0.14, \quad \alpha_{2}=0.12, \quad a_{11}=0.16,  \quad a_{22}=0.1225,\\
&&\text{(a) positive correlated:}  \quad  a_{12}=a_{21}=0.028 ,\\
&&\text{(b) negative correlated:}  \quad  a_{12}=a_{21}=-0.028,
\end{eqnarray*}and the investment period is set to be one year ($T=1$). In this case, we investigate the optimal trading strategy for a risk averse investor who can access two positively or negatively correlated stocks as well as a risk-free asset. We use the time step $h=0.01$, uniform grid with length $\Delta y = (0.01, 0.01)$ in each dimension, and the number of simulated sample paths $M=10^{5}$.

\begin{figure}[!t]
	\subfigure[$\Theta^{2}$ region and trading strategy]{\label{fig:a1}\includegraphics[width=0.52\linewidth]{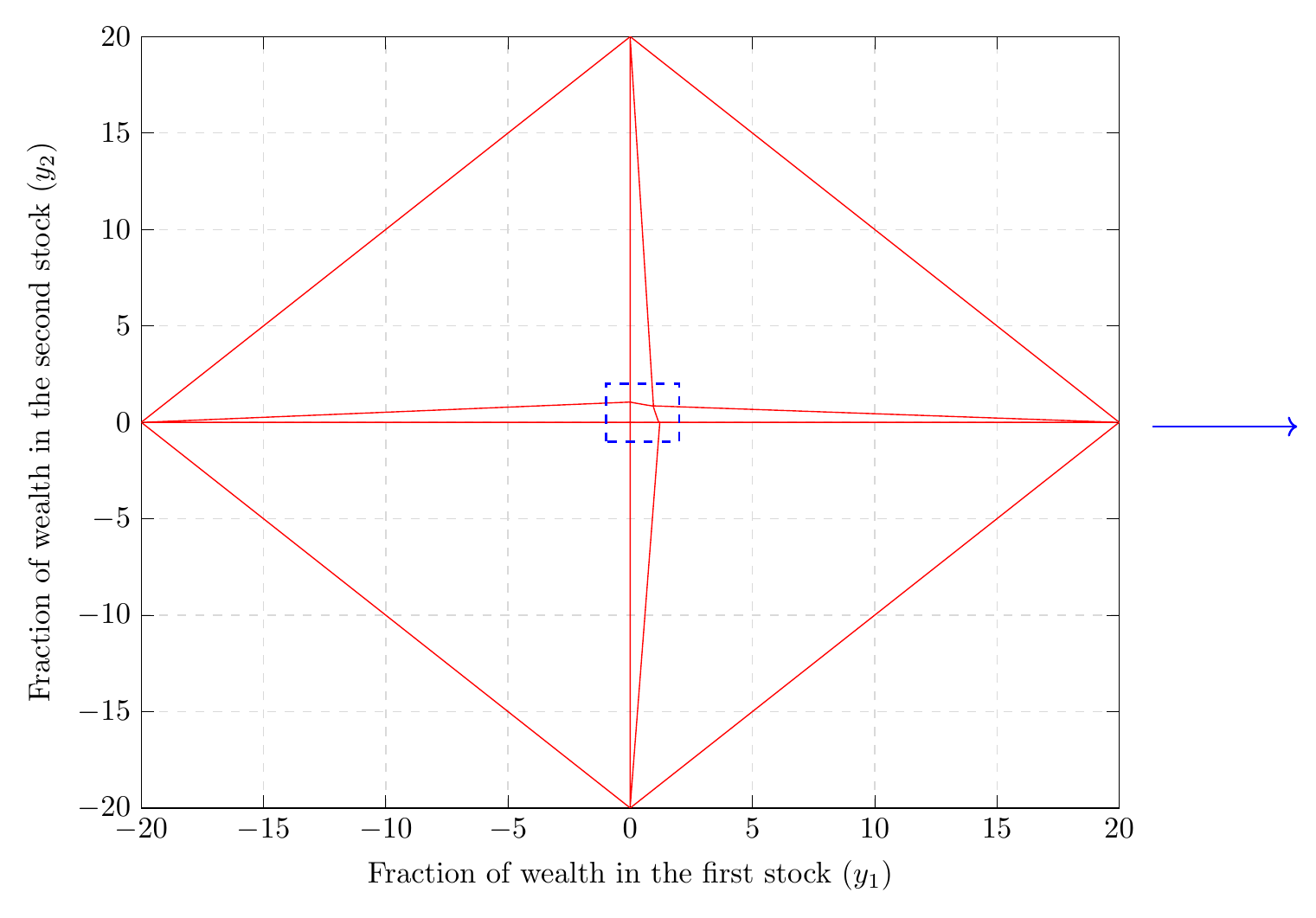}}
	\subfigure[Enlarged no-trading region]{\label{fig:b1}\includegraphics[width=0.46\linewidth]{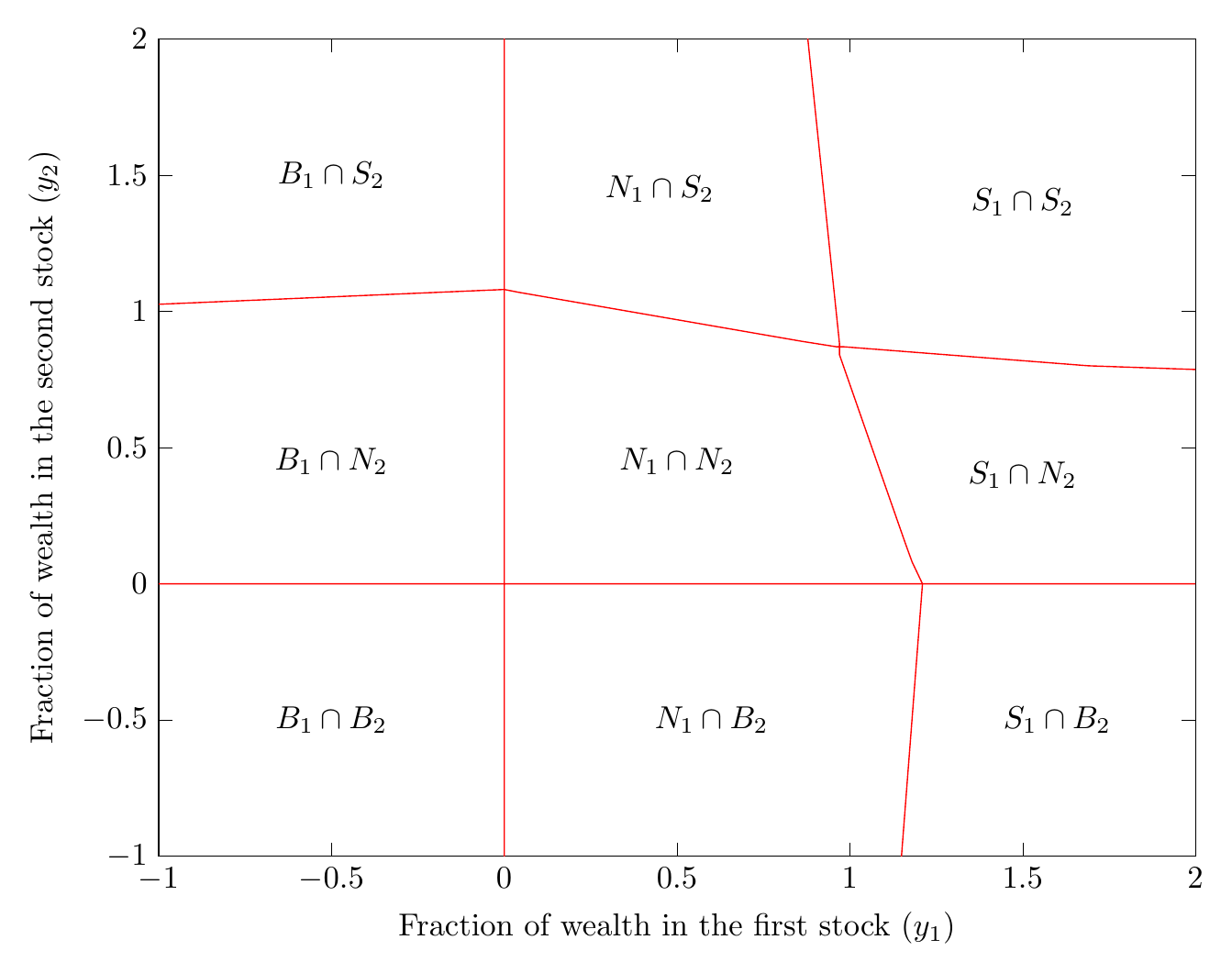}}
	
	\caption{The estimated selling, buying, and no-trading regions for the $N=2$ case at time $t=0.9$. Test 2 parameters: $ \; r=0, \; \alpha_{1}=0.14, \; \alpha_{2}=0.12, \; a_{11}=0.16, \; a_{22}=0.1225, \; a_{12}=a_{21}=0.028, \; \mu_{1}=\lambda_{1}=\mu_{2}=\lambda_{2}=0.05, \; \gamma=0.2, \; \beta=0.1$}
	\label{fig:case}
\end{figure}

Figure \ref{fig:case} shows the decomposition of the domain $\Theta^{2}$ into selling ($S_{i}, i=1, 2$), buying ($B_{i}, i=1, 2$), and no-trading ($N_{i}, i=1, 2$) regions at time $t=0.9$ for the two positively correlated stocks case. It can be observed that the domain $\Theta^{2}$ is partitioned into nine different regions, with the no-trading region $N_{1} \cap N_{2}$ in the center surrounded by trading regions $S_{1} \cap S_{2}$, $S_{1} \cap N_{2}$, $S_{1} \cap B_{2}$, $N_{1} \cap B_{2}$, $B_{1} \cap B_{2}$, $B_{1} \cap N_{2}$, and $B_{1} \cap S_{2}$ in the clockwise order. In addition, the four intersections $\partial S_{1} \cap \partial S_{2}$, $\partial S_{1} \cap \partial B_{2}$, $\partial B_{1} \cap \partial S_{2}$, and $\partial B_{1} \cap \partial B_{2}$ are a singleton. This means that if the initial portfolio position is in $B_{1} \cap S_{2}$, for example, the investor should buy the first stock and sell the second one to reach the unique corner $\partial B_{1} \cap \partial S_{2}$. The phenomena we observed are consistent with rigorous analysis results proved in \cite{chen2013characterization}.

\begin{figure}[!t]
	\subfigure[Positive correlated: $a_{12}=a_{21}=0.028$]{\label{fig:positive}\includegraphics[width=0.48\linewidth]{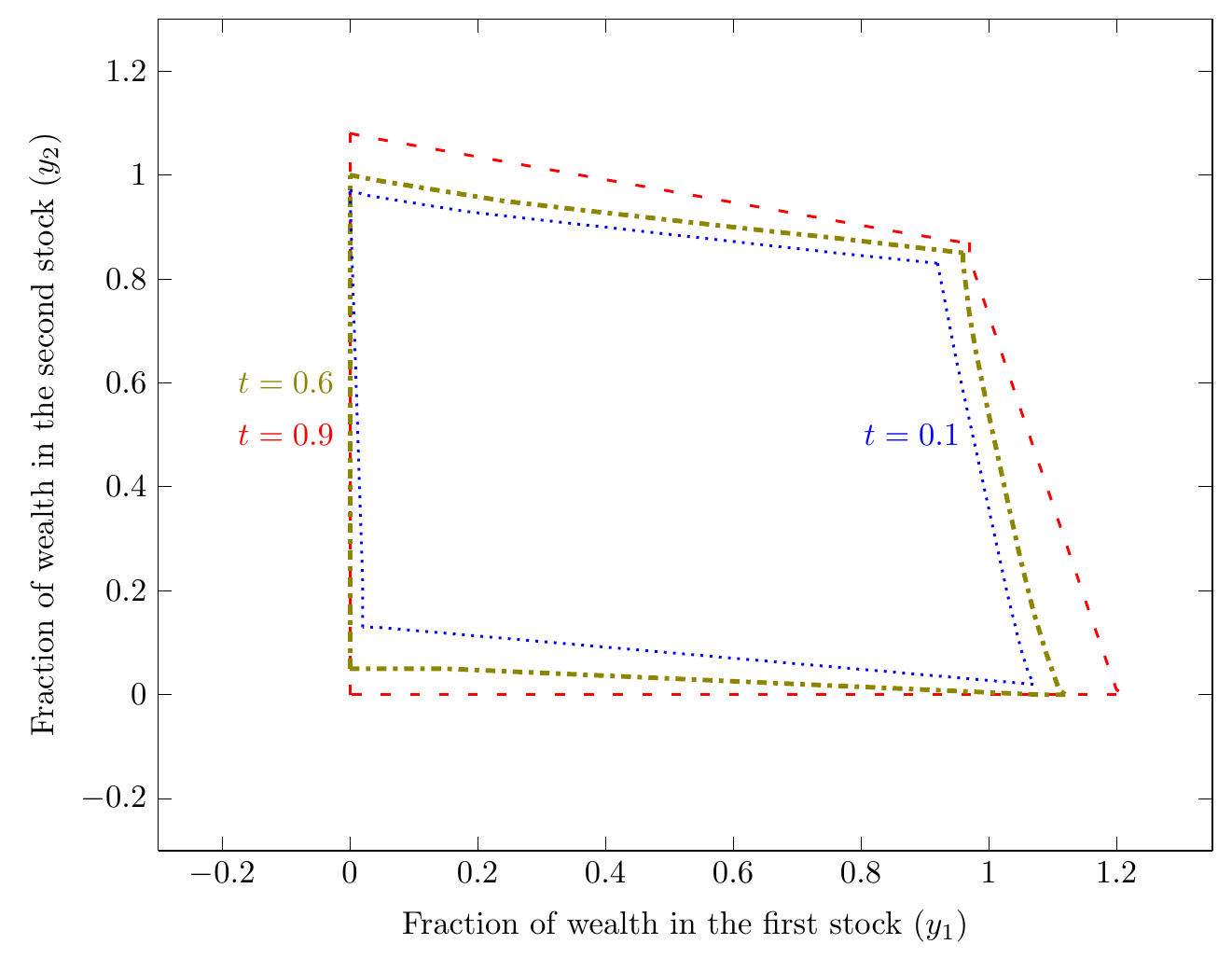}}
	\hfill
	\subfigure[Negative correlated: $a_{12}=a_{21}=-0.028$ ]{\label{fig:negative}\includegraphics[width=0.48\linewidth]{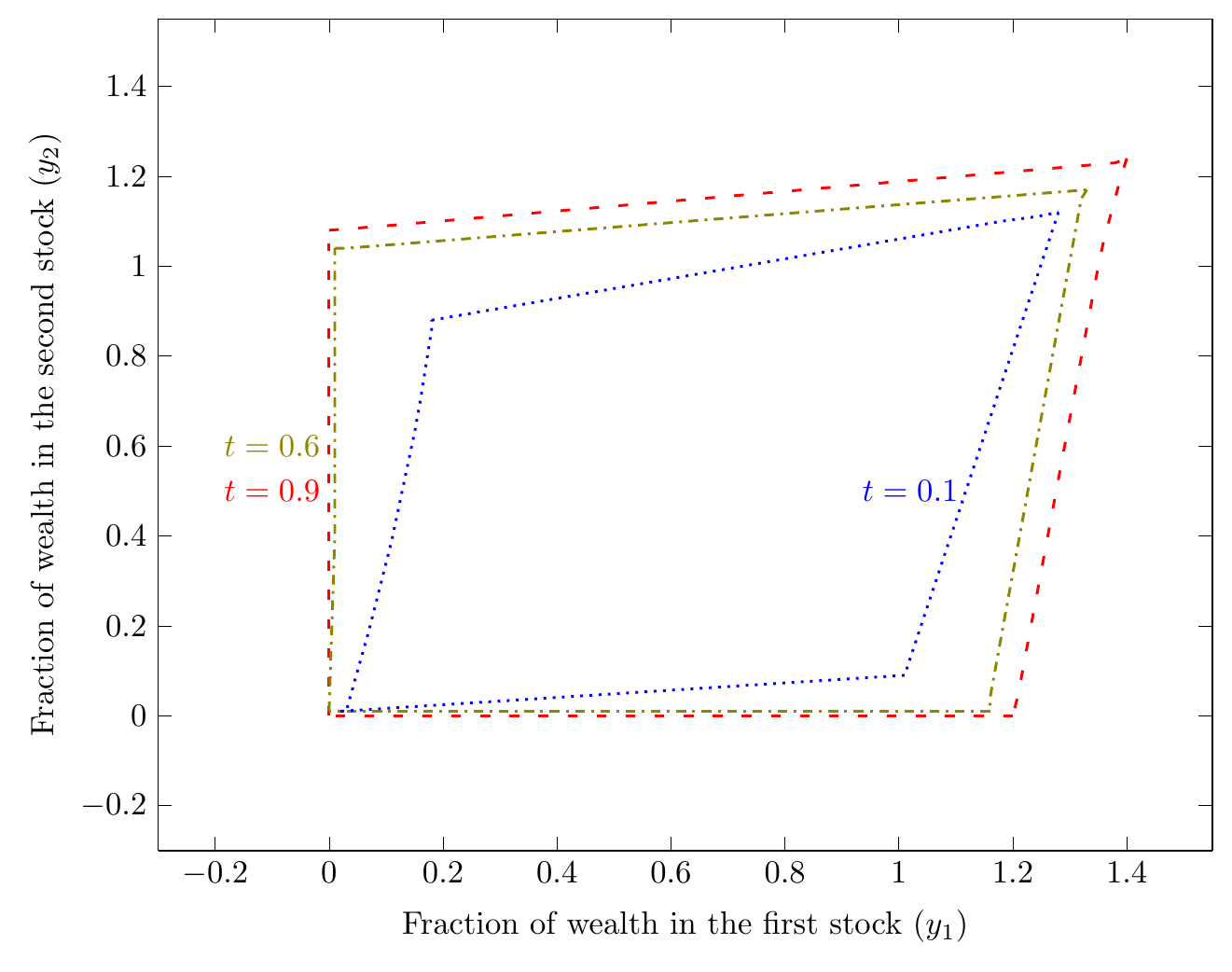}}
	
	\caption{The estimated no-trading region for the $N=2$ case at different time steps. Test 2 parameters:  $ \; r=0, \; \alpha_{1}=0.14, \; \alpha_{2}=0.12, \; a_{11}=0.16, \; a_{22}=0.1225, \; \mu_{1}=\lambda_{1}=\mu_{2}=\lambda_{2}=0.05, \; \gamma=0.2, \; \beta=0.1$}
	\label{fig:image8}
\end{figure}

The numerical approximation of the no-trading region at different time steps is provided in Figure \ref{fig:image8} for both the two positively and negatively correlated stocks cases. Here the expected rate of return for the first stock $\alpha_{1}=14\%$ is more than that of the second stock $\alpha_{2}=12\%$ and transaction costs for buying and selling stocks are kept equal. Since the first stock gives a higher rate of return, as expected the investor will not only put more fraction in the first stock but have a larger inhibition to trade the first one. In Figure \ref{fig:positive} since these two stocks are positive correlated, the region of inaction can only elongate along the main diagonal. An explanation of this behavior is that the investor does not loose much by having more fraction in one stock and less in the other because one partially hedges the other. On the other hand, the result for the two negatively correlated stocks case is displayed in Figure \ref{fig:negative}. As we can see from the figure, the no-trading region elongates along the anti-diagonal direction. This implies that when the price of one performs worse than usual, the other will likely do better than usual. The gain in one stock is therefore likely to offset the loss in the other. Hence, the investor does not loose much by having more fraction in both stocks. These observations are the same as the results obtained in \cite{muthuraman2006multidimensional} for the infinite time horizon problem. Moreover, we can observe that $B_{1}=0$ and $B_{2}=0$ when the time approaches to the maturity of the investment period, which confirms the ``no-buying near maturity'' phenomenon.

\begin{remark} \cite{muthuraman2006multidimensional} provide a computational method to solve the portfolio optimization problem with infinite horizon. They use an iterative scheme that adapts the finite element method in order to capture the region of inaction (please see \cite{muthuraman2006multidimensional} for more detail). Their problem does not depend on time, so it can be focused on finding the no-trading region only. However, if we consider the same optimization problem with finite time horizon, different trading regions should be characterized in order to adjust the value function based on the different regions. 

We notice that if we follow the numerical scheme proposed by \cite{muthuraman2006multidimensional} and adapt the implicit finite difference method instead, we obtain the same no trading region. However, the estimated buying and selling regions are not acceptable. Take $N=2$ for example. Figure \ref{fig:image10} shows the comparison of results at time $t=0.9$ obtained by the mixed Monte Carlo simulation and finite difference method we proposed and the iterative scheme adapting the implicit finite difference method using the same parameter settings in Test 2(a). Observe that when the iterative scheme adapting finite difference method is applied, the numerical result illustrates $B_{1} \neq 0$ and $B_{2} \neq 0$ as the time approaches maturity, which obviously violates the ``no-buying near maturity'' phenomenon. The implementation of the proposed Monte Carlo scheme could cure this problem, and therefore gives a compatible result with the theoretical analysis.  
\begin{figure}[!t]
	\subfigure[Mixed MC and FD Algorithm]{\label{fig:MC}\includegraphics[width=0.48\linewidth]{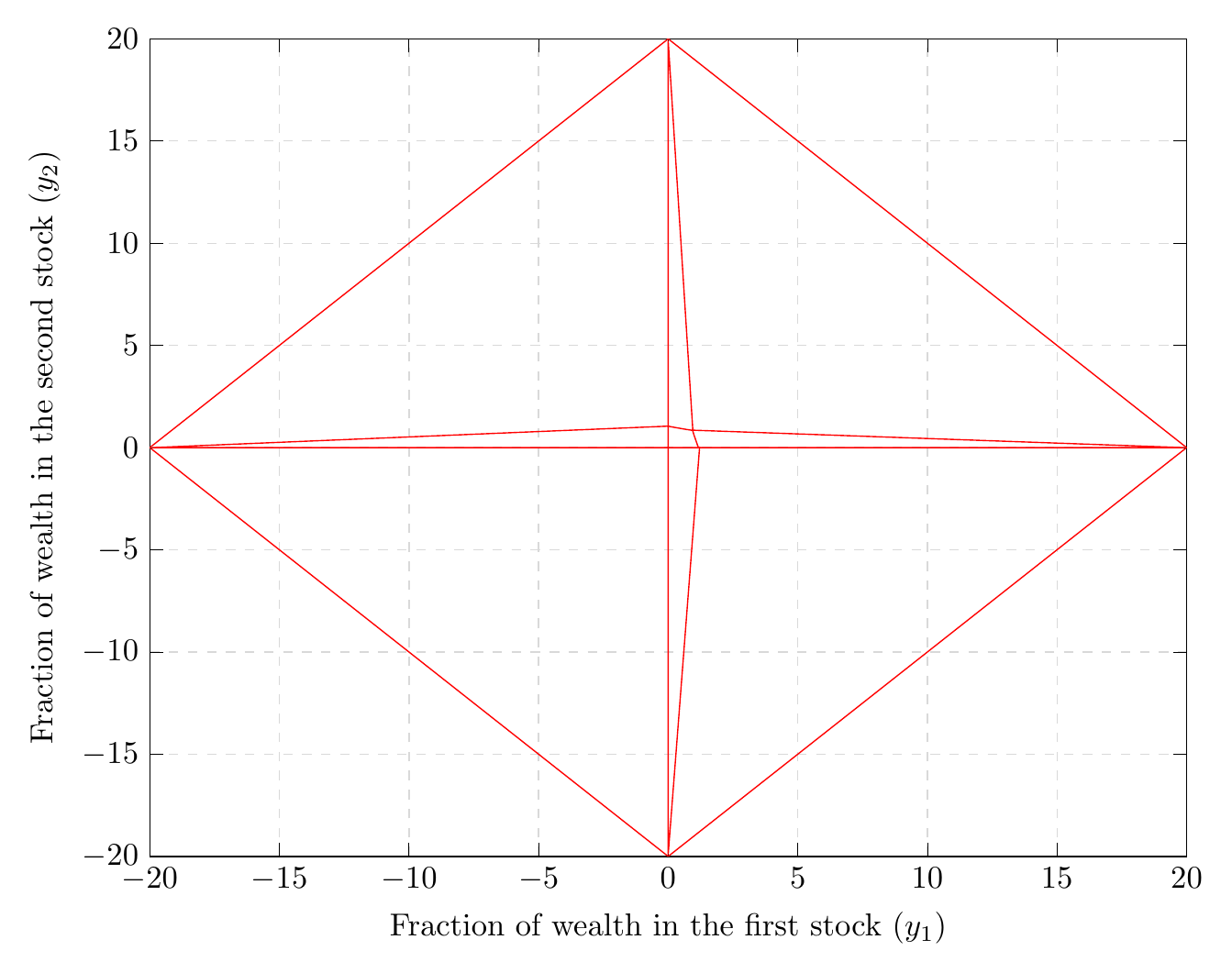}}
	\hfill
	\subfigure[Implicit FD Method]{\label{fig:FD}\includegraphics[width=0.48\linewidth]{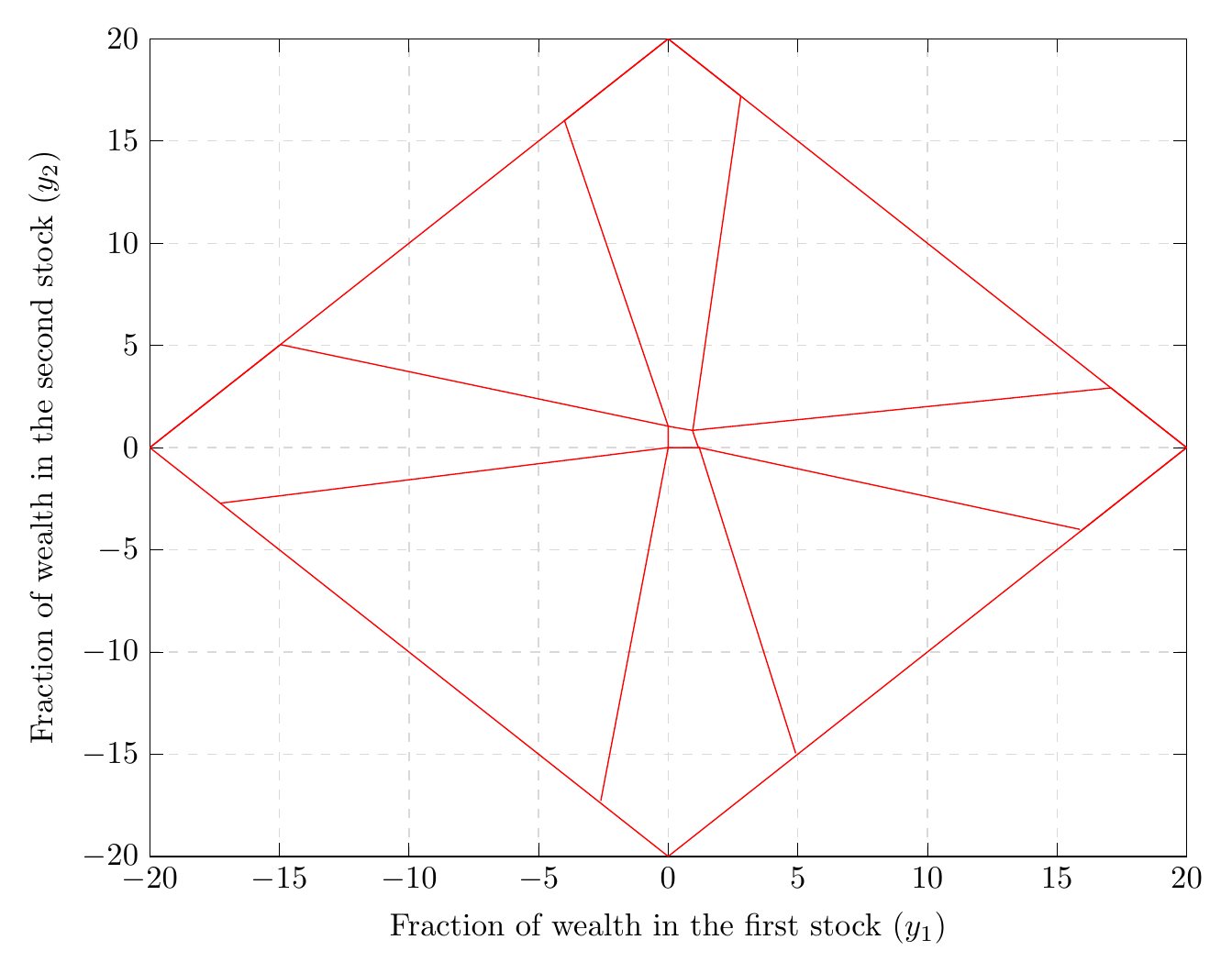}}
	
	\caption{The estimated selling, buying, and no-trading regions for the $N=2$ case at time $t=0.9$. Test 2 parameters: $ \; r=0, \; \alpha_{1}=0.14, \; \alpha_{2}=0.12, \; a_{11}=0.16, \; a_{22}=0.1225, \; a_{12}=a_{21}=0.028, \; \mu_{1}=\lambda_{1}=\mu_{2}=\lambda_{2}=0.05, \; \gamma=0.2, \; \beta=0.1$}
	\label{fig:image10}
\end{figure}
\end{remark}

\subsection{Test 3}
In this numerical test, we consider both the correlated and uncorrelated stocks cases with the following financial parameter values:
\begin{eqnarray*}
&& N=3, \quad r=0.07, \quad \beta=0.1, \quad \gamma=0.2, \quad \mu_{1}=\lambda_{1}=\mu_{2}=\lambda_{2}=\mu_{3}=\lambda_{3}=0.1,\\ 
&& \alpha_{1}=0.14, \quad\alpha_{2}=0.12, \quad \alpha_{3}=0.1, \quad a_{11}=0.16,  \quad a_{22}=0.1225,  \quad a_{33}=0.09,\\ 
&&\text{(a) uncorrelated:}  \quad a_{ij}=0, \; \text{ for } i\neq j ,\\
&&\text{(b) correlated:}  \quad \quad  a_{12}=a_{21}=0.014, \quad  a_{23}=a_{32}=0.0105, \quad  a_{13}=a_{31}=0.012,
\end{eqnarray*}
and the investment period is set to be one year ($T=1$). The numerical result is obtained by using time step $h=0.01$, uniform grid with length $\Delta y = (0.01, 0.01, 0.01)$ in each dimension, and the number of simulated sample paths $M=10^{5}$. We test the computational method for three stocks case for two reasons. First, we would like to demonstrate that the proposed numerical method can be applied to high-dimensional problem. Second, it allows us to see if the insights we have in the two stocks case carry over to higher dimensions.

\begin{figure}[!t]
	\subfigure[Uncorrelated]{\label{fig:noncorr}\includegraphics[width=0.48\linewidth]{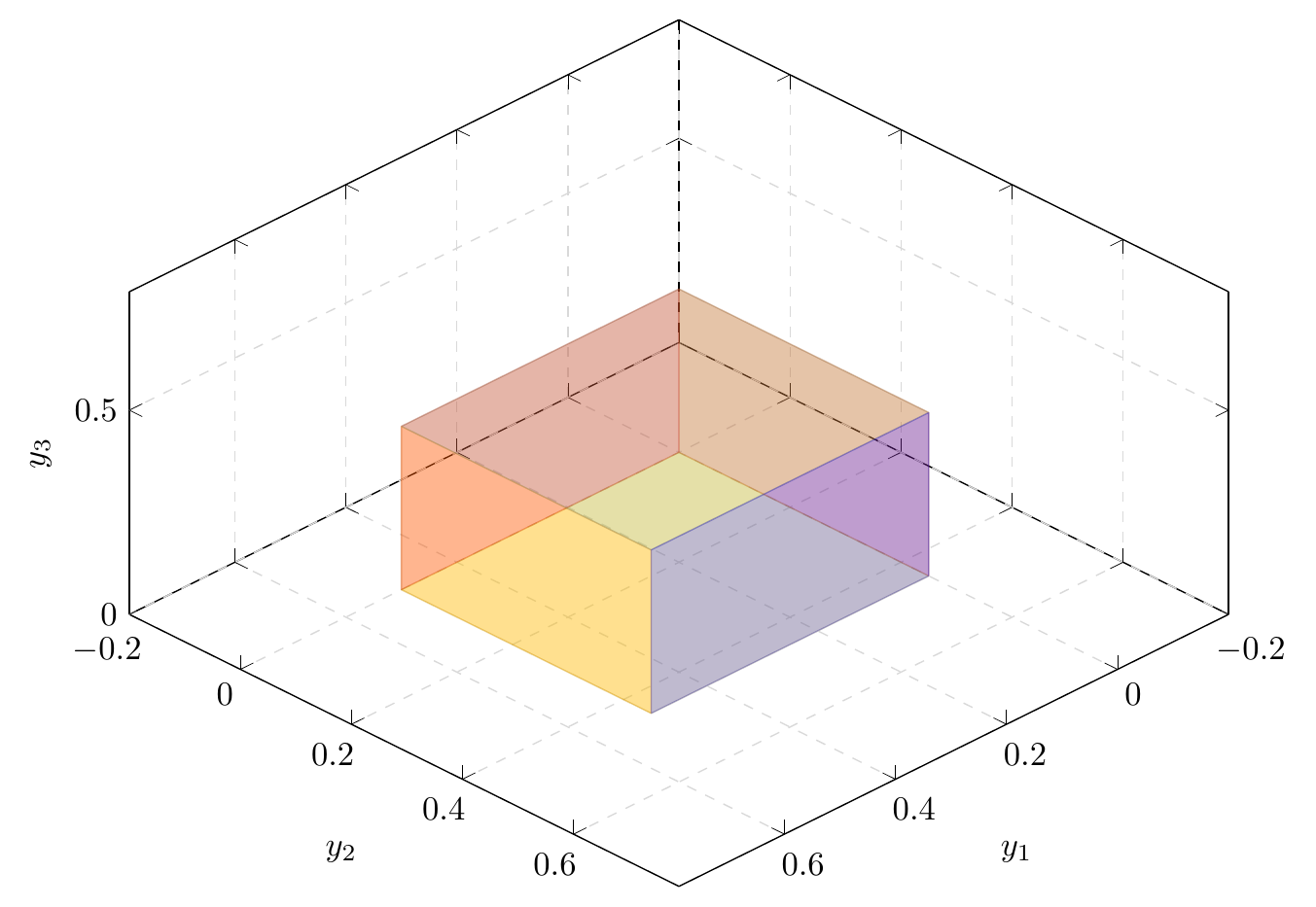}}
	\hfill
	\subfigure[Correlated]{\label{fig:corr}\includegraphics[width=0.48\linewidth]{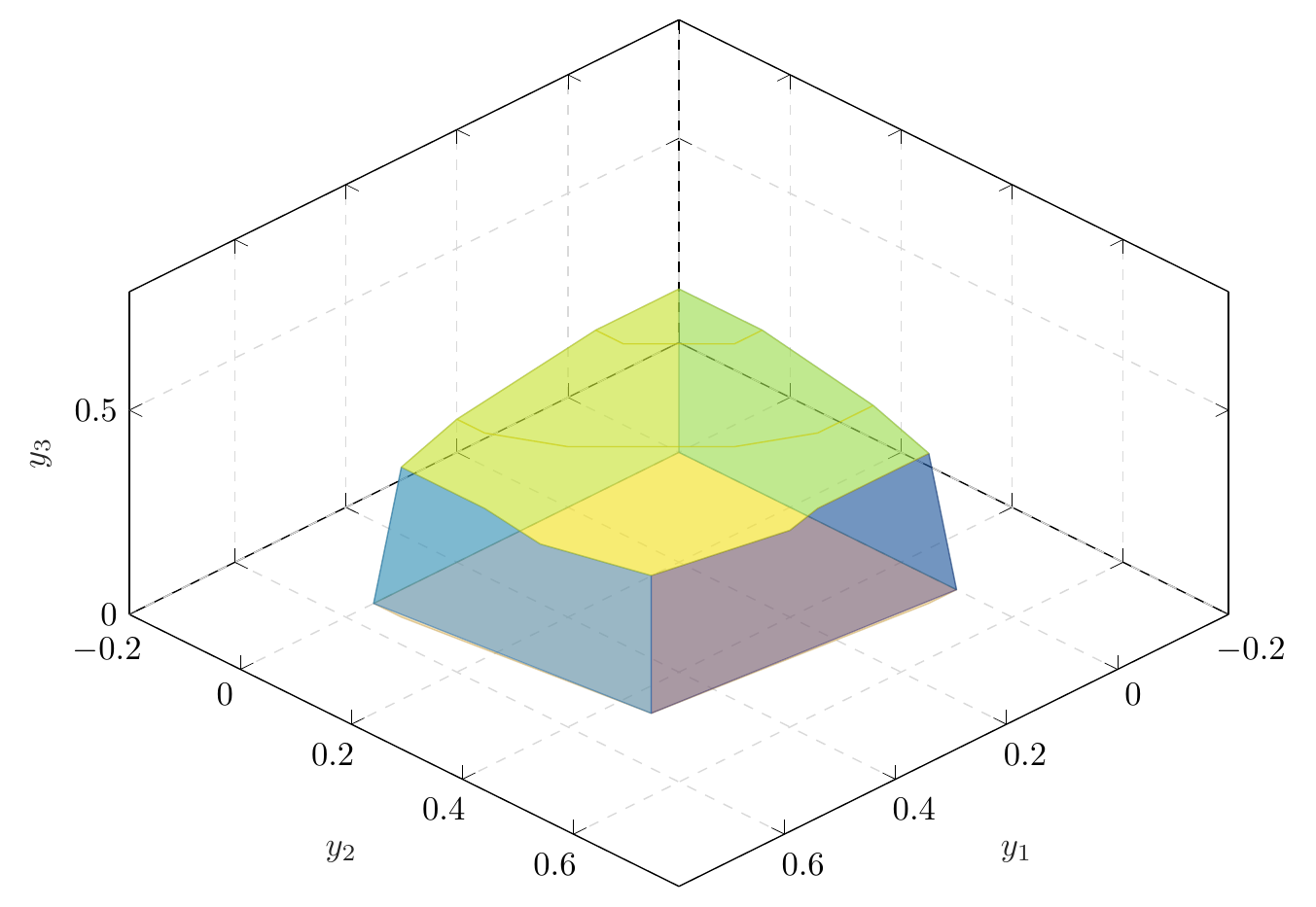}}
	
	\caption{The estimated no-trading region for the $N=3$ case. Test 3 parameters: $ r=0.07, \; \beta=0.1, \; \gamma=0.2, \; \mu_{1}=\lambda_{1}=\mu_{2}=\lambda_{2}=\mu_{3}=\lambda_{3}=0.1, \; \alpha_{1}=0.14, \; \alpha_{2}=0.12, \; \alpha_{3}=0.1,\; a_{11}=0.16,  \; a_{22}=0.1225, \; a_{33}=0.09$. (a) Uncorrelated: $a_{ij}=0, i\neq j$. (b) Correlated: $a_{12}=a_{21}=0.014, a_{23}=a_{32}=0.0105, a_{13}=a_{31}=0.012$. }
	\label{fig:image7}
\end{figure}

Figure \ref{fig:image7} shows the approximated no-trading region for both the three independent stocks case and the three correlated stocks case at time $t=0.9$. First observe that the no-trading region is a closed region set bounded by six surfaces in three dimensions. Also note that the no-trading region of the independent stocks case in Figure \ref{fig:noncorr} is close to a rectangular cubic while the no-trading region of the correlated stocks case in Figure \ref{fig:corr} is askew, which are consistent with previous observations in the two dimensional case. Since the rate of return for the first stock $\alpha_{1}=0.14$ is greater than the other two, a large fraction of wealth in the first stock can be expected, and it is obvious that fewer transactions will be made for the first one. Finally, it also verifies the ``no-buying near maturity'' phenomenon as the time approaches the end of investment period for both cases.

\subsection{Test 4}
In this numerical test, we consider the following financial parameter values:
\begin{eqnarray*}
	&& N=10, \quad r=0.07, \quad \beta=0.1, \quad \gamma=-1, \quad \mu_{i}=\lambda_{i}=10^{-6} \;\; \text{ for }\;\; i=1,\cdots,10,\\ 
	&& \alpha=(0.14,\; 0.14, \; 0.12, \; 0.12, \; 0.1, \; 0.1, \; 0.08, \; 0.08, \; 0.08, \; 0.08)',\\
	&& a = diag( 0.16, \; 0.16, \; 0.1225, \; 0.1225, \; 0.09, \; 0.09, \; 0.0625, \; 0.0625, \; 0.6025, \; 0.0625),\\
	&& a_{ij}=0, \; \text{ for } i\neq j .
\end{eqnarray*}
and the investment period is set to be one year ($T=1$). The numerical result is obtained by using time step $h=0.01$, uniform grid with length $\Delta y_{i} = 0.01$ for $i=1,\cdots,10$ in each dimension, and the number of simulated sample paths $M=10^{5}$. Since transaction costs are really small, this problem is approximately reduced to the Merton's problem. As we can expect, the no-trading region under these parameter settings should be a bounded small region including the \textit{Merton proportion} at any time step. Denote the \textit{Merton proportion} with the power utility function by $\pi^{*}$, and then we have
\begin{equation*}
\pi ^{*} = \frac{1}{(1-\gamma)}a^{-1} (\alpha - re)
\label{powerstrategy multi-asset1} 
\end{equation*}
shown in \cite{merton71}. This solution gives a valuable comparison, and can be used as a benchmark to test whether the algorithm we proposed could provide a qualitative result or not. 

Table \ref{tab:table1} shows the approximated no-trading region for the ten independent stocks case at time $t=0.9$. The lower and upper bounds mean boundaries of the no-trading region in each dimension. We can observe from the table that the no-trading region is such a small region that it is almost the \textit{Merton proportion} point because small transaction costs are applied. This mainly indicates that when the transaction costs are really small, the investor is willing to rebalance his portfolio position so that the proportion of wealth in risky assets is nearly a constant. This result again demonstrates that the proposed numerical method can be applied to high-dimensional problems.

\begin{table}[!t]
\caption{The estimated no-trading region for the $N=10$ case at time $t=0.9$.}
\centering
\begin{tabular}{|c|c|c|c|c|c|c|c|c|c|c|}
	\hline 
	No-trading Region & $y_{1}$ & $y_{2}$ & $y_{3}$ & $y_{4}$ & $y_{5}$ & $y_{6}$ & $y_{7}$ & $y_{8}$ & $y_{9}$ & $y_{10}$ \\ 
	\hline 
	\hline
	Lower Bound & $0.21$ & $0.21$ & $0.20$ & $0.20$ & $0.16$ & $0.16$ & $0.08$ & $0.08$ & $0.08$ & $0.08$ \\ 
	\hline 
	Upper Bound & $0.22$ & $0.22$ & $0.21$ & $0.21$ & $0.17$ & $0.17$ & $0.08$ & $0.08$ & $0.08$ & $0.08$ \\ 
	\hline 
	\hline
	Merton Proportion & $0.2188$ & $0.2188$ & $0.2041$ & $0.2041$ & $0.1667$ & $0.1667$ & $0.08$ & $0.08$ & $0.08$ & $0.08$\\ 
	\hline 
\end{tabular}
\bigskip
\caption*{\footnotesize{
		Test 4 parameters: $N=10, \; r=0.07, \; \beta=0.1, \; \gamma=-1, \; \mu_{i}=\lambda_{i}=10^{-6} \text{ for } i=1,\cdots,10,$ \\ $\alpha=(0.14,\; 0.14, \; 0.12, \; 0.12, \; 0.1, \; 0.1, \; 0.08, \; 0.08, \; 0.08, \; 0.08)', \; $\\ $a = {\rm\bf diag}( 0.16, \; 0.16, \; 0.1225, \; 0.1225, \; 0.09, \; 0.09, \; 0.0625, \; 0.0625, \; 0.6025, \; 0.0625),$ \\ $a_{ij}=0, \; \text{ for } i\neq j.$}}
\label{tab:table1}
\end{table}

\section{Conclusion}
In this paper, we have proposed a mixed numerical method including Monte Carlo simulation and finite difference method to cope with the different difficulties associated with the optimal investment and consumption problem in the presence of transaction costs during a finite investment period, for which no analytical solution exists. The computed approximations satisfy all the qualitative properties which have been theoretically proved for the one risky asset case. Furthermore, the numerical solutions provide the optimal approximated value function in the presence of transaction costs and also determine the behaviors of optimal no-trading, selling, and buying regions.

It is worthwhile to point out that we not only characterize boundaries of the optimal trading policies but also provide admissible heuristics for a portfolio which includes many stocks. We believe the motivation behind this proposed numerical scheme in section 4 can be extended to various HJB models for singular control problems. For instance, it can be directly adapted to the optimal investment problem with transaction costs. 

Indeed, this work carries out many directions of future research. Arguably we do not work on different choices of diffusion coefficients in the Monte Carlo step. On the other hand, in order to obtain a more accurate approximation, we observe that the high-level refinement of the meshes is required as the dimension increases which leads to an increase in computational time. It is important to mention that it does not automatically give a better result if we only refine the grids near the no-trading region and regions in which we trade only one stock. We eventually hope to include theoretical analysis and improved algorithm for these parts.

\bibliographystyle{apalike}	
\bibliography{transaction}
\end{document}